\documentclass[twocolumn,tighten]{aastex631}

\usepackage{ulem}
\usepackage{multirow}

\shortauthors{Fukui et al.}
\shorttitle{The Gamma Ray Origin in RXJ0852}

\begin{document}
\title{The Gamma Ray Origin in RX~J0852.0$-$4622 Quantifying the Hadronic and Leptonic Components: Further Evidence for the Cosmic Ray Acceleration in Young Shell-type SNRs}

\correspondingauthor{Yasuo Fukui}
\email{fukui@a.phys.nagoya-u.ac.jp}

\author[0000-0002-8966-9856]{Yasuo Fukui}
\affiliation{Department of Physics, Nagoya University, Furo-cho, Chikusa-ku, Nagoya 464-8601, Japan}

\author[0000-0001-5069-5988]{Maki Aruga}
\affiliation{Department of Physics, Nagoya University, Furo-cho, Chikusa-ku, Nagoya 464-8601, Japan}

\author[0000-0003-2062-5692]{Hidetoshi Sano}
\affiliation{Faculty of Engineering, Gifu University, 1-1 Yanagido, Gifu 501-1193, Japan}

\author[0000-0003-0324-1689]{Takahiro Hayakawa}
\affiliation{Department of Physics, Nagoya University, Furo-cho, Chikusa-ku, Nagoya 464-8601, Japan}

\author[0000-0002-7935-8771]{Tsuyoshi Inoue}
\affiliation{Department of Physics, Faculty of Science and Engineering, Konan University, 8-9-1 Okamoto, Higashinada, Kobe, Hyogo 658-8501, Japan}

\author[0000-0002-9516-1581]{Gavin Rowell}
\affiliation{School of Physical Sciences, The University of Adelaide, North Terrace, Adelaide, SA 5005, Australia}

\author[0000-0001-9687-8237]{Sabrina Einecke}
\affiliation{School of Physical Sciences, The University of Adelaide, North Terrace, Adelaide, SA 5005, Australia}

\author[0000-0002-1411-5410]{Kengo Tachihara}
\affiliation{Department of Physics, Nagoya University, Furo-cho, Chikusa-ku, Nagoya 464-8601, Japan}

\begin{abstract}
\citet{2021ApJ...915...84F} quantified the hadronic and leptonic gamma rays in the young TeV gamma ray shell-type supernova remnant (SNR) RX~J1713.7$-$3946 (RXJ1713), and demonstrated that the gamma rays are a combination of the hadronic and leptonic gamma ray components with a ratio of $\sim 6:4$ in gamma ray counts $N_\mathrm{g}$.
This discovery, which adopted a new methodology of multiple-linear gamma-ray decomposition, was the first quantification of the two gamma ray components.
In the present work, we applied the same methodology to another TeV gamma ray shell-type SNR RX~J0852.0$-$4622 (RXJ0852) in the 3D space characterized by [the interstellar proton column density $N_{\mathrm{p}}$]-[the nonthermal X-ray count $N_{\mathrm{x}}$]-[$N_{\mathrm{g}}$], and quantified the hadronic and leptonic gamma ray components to have a ratio of $\sim 5:5$ in $N_{\mathrm{g}}$.
The present work adopted fitting of two/three flat planes in the 3D space instead of a single flat plane, which allowed to suppress fitting errors.
The quantification indicates that the hadronic and leptonic gamma rays are in the same order of magnitude in these two core-collapse SNRs, verifying the significant hadronic gamma ray components.
We argue that the target interstellar protons, in particular their spatial distribution, are essential in any attempts to identify type of particles responsible for the gamma-ray emission.
The present results confirm that the CR energy $\lesssim 100$\,TeV is compatible with a scheme that SNRs are the dominant source of these Galactic CRs.
\end{abstract}


\defcitealias{2012ApJ...746...82F}{Paper I}
\defcitealias{2017ApJ...850...71F}{Paper II}
\defcitealias{2021ApJ...915...84F}{Paper III}
\defcitealias{2010ApJ...708..965Z}{ZA10}

\section{Introduction}\label{sec:introduction}
\subsection{Gamma Ray Production in the Galaxy}\label{subsec:introduction:gamma_ray_production}
The origin of the Galactic cosmic rays (CRs) has been a long-standing issue in astrophysics.
For cosmic rays of energy less than the knee at $<10^{15.5}$\,eV, supernova remnants (SNRs) are the most promising candidate for CR acceleration in the Galaxy.
The diffusive shock acceleration provides an effective mechanism to accelerate charged particles which account for the acceleration of CRs in SNRs \citep{1978MNRAS.182..147B}.
In order to verify the CR acceleration in SNRs, it is essential to identify gamma rays and/or neutrinos, both of which follow the p-p reactions of the CR protons.
The observations of SNRs with the gamma ray instruments including H.E.S.S., MAGIC, VERITAS, \textit{Fermi}-LAT, etc.\ detected gamma rays toward a number of SNRs in the last two decades.
The neutrino imaging is being developed, while it is not yet at a stage comparable to the gamma ray imaging \citep[e.g.,][]{2019arXiv190102528M}.
Most recentry IceCube Collaboration revealed strong concentration of neutrino emission in the Galactic plane, lending support for the SNR origin of the neutrinos \citep{2023Sci...380.1338I}.
The very high energy gamma rays in the GeV-TeV range are therefore the current best probe for the CRs.
In particular, the young TeV gamma ray SNRs with ages of 1000--5000\,yrs are most important because their gamma ray energy range is the highest at $\gtrsim 100$\,TeV in the Galaxy, reaching fairly close to the knee, which is likely linked with the highest energy CRs in the Galaxy.
These SNRs include RX~J1713.7$-$3946 (RXJ1713), the brightest one in the sky, and RX~J0852.0$-$4622 (RXJ0852) as the second brightest.
This contrasts with the middle aged SNRs having lower gamma ray energy by three orders of magnitude.
Middle aged SNRs do not exhibit ongoing acceleration of the highest energy CRs, but their escaping CRs are important tracers of past CR acceleration \citep[e.g.,][]{2021MNRAS.503.3522M,2010PASJ...62.1127C}.

\subsection{The Two Gamma Ray Components}\label{subsec:introduction:two_gamma_ray_components}
The CRs create gamma rays via two processes, i.e., the hadronic process ($\mathrm{pp}\rightarrow \pi^{0}~\rightarrow 2\gamma$) and the leptonic process (Inverse Compton (IC) scattering).
Quantification of the CR energy will be accomplished only by identifying both the hadronic gamma ray components and the interstellar protons, the target for the p-p reactions.
Otherwise, one cannot exclude a possibility that the gamma rays are dominated by the leptonic gamma ray components that carry only the minor part of the CR energy.
It was thought that the gamma ray spectrum is able to offer a key signature which can discriminate the two components.
This, however, turned out to be hardly achieved, because the gamma ray spectrum is dramatically changeable not only by the process but also by the properties of the CRs and the interstellar medium (ISM) in SNRs.
For example, \citet{2011ApJ...734...28A} claimed that the gamma rays obtained by Fermi collaboration in RXJ1713 are due to the leptonic process by presenting a hard GeV--TeV gamma ray spectrum similar to what is expected from the leptonic process.
On the other hand, \citet{2012ApJ...744...71I} argued that density-dependent penetration of CR protons into dense cloud cores produces a hadronic gamma ray spectrum very similar to the leptonic gamma ray spectrum.
Such dense cloud cores are in fact observed in the gamma ray peaks in RX J1713 by \citet{2003PASJ...55L..61F}, \citet[][Paper I hereafter]{2012ApJ...746...82F}, \citet{2005ApJ...631..947M} and \citet{2012MNRAS.422.2230M,2013PASA...30...55M}, and may significantly affect the gamma ray spectrum.
The hard gamma ray spectrum compared with the GeV-TeV gamma ray spectrum was calculated for uniform low-density target protons \citep{2011ApJ...734...28A}, which is not realistic.
\citet{2012ApJ...744...71I} therefore suggested that only spatial correspondence between the interstellar protons and the gamma rays, instead of the gamma ray spectrum, can verify the hadronic gamma rays.
A similar argument on the spectrum was presented in terms of the cosmic ray diffusion in the diffuse gamma rays by \citet{2007ApJ...665L.131G}, and such a modified hard hadronic spectrum was confirmed through numerical simulations by \citet{2014MNRAS.445L..70G}. \citet{2018A&A...612A...6H} and references therein \citep[see also][]{2007A&A...464..235A} repeatedly showed that the observed gamma ray spectrum of RXJ1713 can be fit well by using either of the pure hadronic and pure leptonic models under reasonable sets of CR parameters, while for simplicity they assumed uniform ISM protons instead of the realistic clumpy ISM distribution \citepalias[see for details below]{2012ApJ...746...82F}.
In summary, the spectral fit is not a tool which can by iteself discriminate the two gamma ray origins.

\subsection{Gamma Ray Imaging and Its Spatial Correspondence}\label{subsec:introduction:gamma_ray_imaging}
The hadronic gamma rays will naturally follow the observed distribution of the target interstellar protons, and the spatial distribution of the gamma rays is an essential piece of the hadronic gamma ray properties.
The early gamma ray imaging was too low in spatial resolution at a degree scale (e.g., EGRET).
The advent of the atmospheric Cerenkov telescopes improved resolution in particular in the TeV range.
Among all, the H.E.S.S.\ obtained TeV gamma ray images with a resolution of $\sim 0\fdg 1$, which can resolve nearby young SNRs with ages of 1000--5000\,yrs having spatial extents of around a degree.
These SNRs include RXJ1713, RXJ0852, HESS~J1731$-$347 (HESSJ1731), RCW~86, and several more in the TeV range \citep[The H.E.S.S.\ Galactic plane survey,][]{2018A&A...612A...1H}.

In RXJ1713, the CANGAROO Cerenkov telescope detected and mapped TeV gamma rays \citep{2002Natur.416..823E}, and \citet{2003PASJ...55L..61F} compared the NANTEN CO observations with the partial TeV gamma ray image.
\citet{2003PASJ...55L..61F} revealed that one of the CO peaks, a dense molecular cloud core in the SNR, shows a good spatial match with the TeV gamma ray peak, and suggested that the coincidence is a possible signature of the hadronic gamma ray components.
Subsequently, H.E.S.S.\ collaboration revealed good spatial correspondence between the TeV gamma ray peaks and the NANTEN CO clouds toward the middle aged SNR W~28, indicating the presence of hadronic gamma rays \citep{2006A&A...449..223A}.
Soon after, in RX J1713 and RXJ0852 H.E.S.S.\ revealed their shell-like TeV gamma ray images, whereas their comparisons with NANTEN CO clouds indicated that the CO did not show overall matching with the gamma ray shells, being unfavorable for the hadronic interpretation \citep{2007A&A...464..235A,2007ApJ...661..236A}.
In particular, it was revealed that strong gamma rays are emitted even in the regions with no CO emission.

\citetalias{2012ApJ...746...82F} \citep[see also][]{2008AIPC.1085..104F} renewed the total interstellar proton distribution in RXJ1713 based on the CO and \ion{H}{1} gas, where the new \ion{H}{1} survey by \citet{2005ApJS..158..178M} at $2\arcmin$ resolution was employed.
A new aspect of \citetalias{2012ApJ...746...82F} was the inclusion of the atomic hydrogen gas having densities of 10--100\,cm$^{-3}$, which was neglected in the previous comparisons under an assumption that CO gas only is important as target protons.
As a result, \citetalias{2012ApJ...746...82F} revealed that the total interstellar protons consisting of both molecular and atomic protons have good spatial correspondence with the gamma rays, indicating that dense \ion{H}{1} gas is equally important with the CO gas toward gamma-ray bright regions.
Further, in RXJ0852 \citet[Paper II hereafter]{2017ApJ...850...71F} carried out a comparative study of the total ISM protons of both atomic and molecular gas with the TeV gamma rays, and found that RXJ0852 also shows good spatial correspondence with the total ISM protons which include both molecular and atomic protons.
\citetalias{2012ApJ...746...82F} and \citetalias{2017ApJ...850...71F} therefore lend support for a hadronic gamma ray component both in RXJ1713 and RXJ0852 under an assumption of uniform CR energy density.

Nonetheless, the spatial correspondence alone between the gamma rays and the target ISM protons does not exclude the contribution of the leptonic components.
In an extreme case in the opposite, suppose for instance that the leptonic gamma rays is overwhelming the hadronic gamma rays everywhere in an SNR, while the distributions of the gamma rays, nonthermal X rays, and the ISM protons all appear shell-like.
In such a case, the gamma ray distribution is determined by the leptonic components, and the correspondence between the gamma rays and the ISM protons is fortuitous.
We therefore need a more accurate quantitative method for verification of the gamma ray origins.
In this context, it helps to consider a case where the two gamma ray components are coexistent in the gamma ray counts.
Such a hybrid picture was suggested in HESSJ1731 and RCW~86, which exhibit that part of the gamma rays shows spatial correspondence with the ISM and the rest with the nonthermal rays \citep{2014ApJ...788...94F,2019ApJ...876...37S}.
In the hybrid case, we expect naively that the gamma ray counts will increase with both the ISM proton column density and the nonthermal X ray counts, which may become obvious in scatter plots of the gamma ray counts vs.\ the ISM proton column density or vs.\ the nonthermal X ray counts.
If the hybrid case is correct, the two SNRs HESSJ1731 and RCW~86 imply that the two gamma ray components maybe similar to each other.
However in these cases the number of pixels available for study is $\leq 10$, limiting the precision of the conclusions.

\subsection{Quantification of the Hadronic and Leptonic Gamma Rays in RXJ1713}\label{subsec:introduction:quantification_of_the_hadronic_and_leptonic_gamma_rays}
Recently, in RXJ1713 H.E.S.S.\ gamma ray data were updated and released after significant improvement by \citet{2018A&A...612A...6H}.
The new TeV gamma ray data ($>1$\,TeV energy range) have a higher angular resolution by a factor of three, 1.4\,pc, along with a factor of two better sensitivity.
\citet[Paper III hereafter]{2021ApJ...915...84F} used the data and applied a new methodology which enabled quantification of the two gamma ray origins for the first time, and derived a hadronic : leptonic gamma ray count ratio of $\sim 6:4$ in RXJ1713.
We note that \citetalias{2021ApJ...915...84F} is distinguished from the previous theoretical results \citep[e.g.,][]{2018A&A...612A...6H} because the ISM proton distribution adopted in \citetalias{2021ApJ...915...84F} is obtained directly from the CO and \ion{H}{1} observations in \citetalias{2012ApJ...746...82F}, which ensures sufficiently high accuracy in calculating the hadronic gamma ray components.
This accuracy cannot be attained in the other works that assume uniform ISM distribution which has no observational justification.

The new methodology of \citetalias{2021ApJ...915...84F} quantifies the leptonic and hadronic gamma ray components as follows; the hadronic gamma ray counts are proportional to the number product of the CR protons and the target interstellar protons in the p-p reactions and the leptonic gamma ray counts are proportional to the number product of the CR electrons and the target low energy photons, usually the cosmic microwave background (CMB) photons, in the IC scattering.
These relationships allow one to formulate the gamma ray counts [$N_{\mathrm{g}}$] as a combination of two linear terms; one is proportional to the X ray counts [$N_{\mathrm{x}}$], as a proxy of the CR electrons, and the other proportional to the ISM proton column density [$N_{\mathrm{p}}$] under an assumption that the CR energy density and magnetic field are uniform.
In RX J1713, the distributions of $N_{\mathrm{p}}$ and $N_{\mathrm{x}}$ are shell-like, similar to the TeV gamma ray distribution $N_{\mathrm{g}}$, whereas, to be strict, distributions of $N_{\mathrm{p}}$ and $N_{\mathrm{x}}$ are spatially different.
The difference is expressed by the spatial correlation coefficient between $N_{\mathrm{p}}$ and $N_{\mathrm{x}}$, which is calculated to be 0.7 using pixel sizes of $0\fdg 1$ \citepalias{2012ApJ...746...82F}.
If the two distributions were completely identical, the coefficient is equal to 1.0.
The difference produces, not vastly, but significantly different spatial distribution of the hadronic and leptonic gamma rays.
Then, \citetalias{2021ApJ...915...84F} applied the formulation to the data pixels of RXJ1713 in a 3D space subtended by $N_{\mathrm{p}}$-$N_{\mathrm{x}}$-$N_{\mathrm{g}}$ and showed that $N_{\mathrm{g}}$ is expressed by a flat tilted plane in the space.
This regression plane is used to derive the hadronic and leptonic gamma ray counts in each pixel, and \citetalias{2021ApJ...915...84F} obtained their ratio to be $\sim 6:4$ in $N_{\mathrm{g}}$ over the whole SNR.
This indicates that the two gamma ray components are comparable with each other.
The next issue is to apply the methodology to the other SNRs and to acquire a broader and deeper view on the CR acceleration.
A larger sample of gamma ray SNRs obtained will allow us to elucidate significant details of the CR acceleration.

\subsection{The TeV Gamma Ray SNR RXJ0852}\label{subsec:introduction:rxj0852}
RX~J0852.0$-$0462 (RXJ0852, G266.2$-$1.2) is an SNR with a large diameter of $2\arcdeg$ at a distance of 750\,pc \citep[e.g.,][]{2008ApJ...678L..35K,2015ApJ...798...82A}.
The SNR has nonthermal X ray emission without thermal features with age of $(2.4\mbox{--}5.1)\times10^{3}$\,yr \citep[][]{2015ApJ...798...82A}.
H.E.S.S.\ collaboration imaged the shell-like TeV gamma ray distribution of RXJ0852, which looks similar to the X ray shell.
These properties are common to RXJ1713.
RXJ0852 will enable testing spatial correspondence between the gamma rays and the ISM thanks to its large apparent size, and is a second promising target where the CR origin can be quantified.
The supernova is a core-collapse type \citep{1998Natur.396..141A,2001ApJ...548L.213M,2001ApJ...559L.131P,2001ApJ...548..814S}.
The associated interstellar gas has been revealed as part of a molecular supershell of $\sim 70$\,pc diameter by NANTEN CO observations (\citetalias{2017ApJ...850...71F}, see also Appendix \ref{sec:associated_ISM}, and for CO supershells \citealt{1999PASJ...51..751F,2001PASJ...53.1003M}).
\citetalias{2017ApJ...850...71F} analyzed the ISM proton distribution toward RXJ0852, which includes both molecular and atomic gas, and showed that the total ISM protons associated with the SNR correspond well spatially to the gamma ray distribution. 

The present paper aims at quantifying the two gamma ray components in RXJ0852 and is organized as follows.
Section \ref{sec:datasets} describes the datasets of CO, \ion{H}{1}, TeV gamma rays and X rays used in the paper, and Section \ref{sec:formulation_and_fitting} presents the formulation and the results of the 3D fitting in the $N_{\mathrm{p}}$-$N_{\mathrm{x}}$-$N_{\mathrm{g}}$ space of RXJ0852 along with supplementary analysis results for RX J1713.
In Section \ref{sec:discussion} we discuss the CR properties shown by the results along with a comparison with RX J1713, and give conclusions in Section \ref{sec:conclusions}.

\section{Observational Data}\label{sec:datasets}
The present work utilized the three datasets of the gamma rays, the X rays, and the interstellar protons as described in the following.

\subsection{RXJ0852}
\subsubsection{HESS TeV Gamma Rays}\label{rxj0852:gamma}
As the gamma ray data we used the excess count map at $E > 100$\,GeV obtained by \citet{2018A&A...612A...7H} as given in Table \ref{tab:summary_multireg}.
The angular resolution is given by a point spread function (PSF) of $0\fdg08$ at 68\% containment radius, which translates to FWHM $\sim 0\fdg 19$ ($\sim 11\farcm 3$).
We adopted a pixel size of $\sim 0\fdg 20$ ($\sim 11\farcm 4$) for pixel-to-pixel comparison, which corresponds to $\sim 2.5$\,pc at a distance of 750\,pc (see also Table \ref{datasets}).
For more details see \citet{2018A&A...612A...7H}.

\begin{deluxetable}{llRR}
\tablecaption{Summary of the TeV Gamma-Ray Data}\label{tab:summary_multireg}
\tablewidth{0pt}
\label{datasets}
\tablehead{
\colhead{Name} & \colhead{Energy band} & \colhead{Pixel size} & \colhead{$n$}}
\decimalcolnumbers
\startdata
RX~J0852.0$-$4622 & $E>100$\,GeV & 11\farcm 4 & 52\\
RX~J1713.7$-$3946 & $E>2$\,TeV & 4\farcm 8 & 75
\enddata
\tablecomments{
Columns (1): source name, (2): energy band, (3): pixel size, (4): number of pixels of the data set (see Sections \ref{rxj0852:gamma} and \ref{rxj1713:data}).}
\end{deluxetable}

\subsubsection{Suzaku X-Rays}
We analyzed \textit{Suzaku} archive data of RX J0852.0-4622 from the Data Archives and Transmission System (DARTS at ISAS/JAXA)\footnote{\url{https://www.darts.isas.jaxa.jp/index.html.en}}.
Details of the data reduction are given below. Table \ref{tab:suzaku} summarizes the information on the 45 pointings in total.
Part of the data was already published by \citet{2016PASJ...68S..10T} and \citet{2017ApJ...850...71F}.

In the present analysis of the image, we employed X-ray Imaging Spectrometer \citep[XIS:][]{2007PASJ...59S..23K}.
XIS consists of four CCD cameras, XIS0, XIS1, XIS2, and XIS3, which were installed at the focal plane of the X-ray telescopes \citep[XRTs:][]{2007PASJ...59S...9S}, where XIS2 was not usable due to the damage by the micrometeorite on November 9 in 2006.
In addition, XIS0 had an anomaly in its segment A on June 23, 2009.
In the present work, we used only the data obtained with XIS0 (excluding segment A after June 23, 2009), XIS1, and XIS3.
These observations utilized spaced-row charge injection \citep[SCI:][]{2008PASJ...60S...1N,2009PASJ...61S...9U} except for observation ID 500010010.

In the present analysis we used ``cleaned event files'' processed and screened by HEASoft version 6.11\footnote{\url{https://heasarc.gsfc.nasa.gov/docs/software/lheasoft/}} and Suzaku pipeline versions 2.0, 2.2, and 2.8.
In the procedure, we made images of photon count at 2.0--5.7\,keV, where non-X-ray background (NXB) was calculated and subtracted from the night time observed data of the earth by using \texttt{xisnxbgen}.
We also made Monte Carlo calculations of the flat-field imaging \citep{2007PASJ...59S.113I} in order to correct for the vignetting effect by XRT.
We here considered the effect of the pixels which were not usable in SCI by using \texttt{xisexpmapgen}.
In the exposure-corrected and background-subtracted count map (normalized in a unit of photons\,s$^{-1}$\,degree$^{-2}$), CCO AX~J0851.9$-$4617.4 $(\alpha_{\mathrm{J2000}}, \delta_{\mathrm{J2000}}) \sim (08^{\mathrm{h}}52^{\mathrm{m}}01\fs 4, -46\arcdeg 17\arcmin53\arcsec)$ \citep{1998Natur.396..141A,1999A&A...350..997A,2001ApJ...548..814S} and the pulsar wind nebula (PWN) $(\alpha_{\mathrm{J2000}}, \delta_{\mathrm{J2000}}) \sim (08^{\mathrm{h}}55^{\mathrm{m}}36\fs 18, -46\arcdeg44\arcmin13.4\arcsec)$ \citep{2013A&A...551A...7A} in the center of the image were masked by a circle of $150\arcsec$ radius.
The resultant image was binned in the same pixels with the H.E.S.S.\ gamma rays, and was used in the present study.

\startlongtable
\begin{deluxetable*}{rlRRcrl}
\tablecaption{Summary of the \textit{Suzaku} XIS Archive Data of RX~J0852.0$-$4622}\label{tab:suzaku}
 \tablehead{
\colhead{No.} & \colhead{ObsID} & \colhead{$\alpha_{\mathrm{J2000}}$} & \colhead{$\delta_{\mathrm{J2000}}$} & \colhead{Start Date} & \colhead{Exposure\tablenotemark{a}} & \colhead{SCI}\\
&& \colhead{($\arcdeg$)} & \colhead{($\arcdeg$)} & \colhead{(yyyy-mm-dd hh:mm)} & \colhead{(ks)} &
}
\startdata
0 & 500010010 & 132.2926 & -45.6157 & 2005-12-19 10:44 & 171 & OFF\\ 
1 & 502023010 & 131.9787 & -45.8064 & 2007-07-04 07:33 & 9 & ON\\ 
2 & 502024010 & 132.1691 & -45.7748 & 2007-07-04 16:34 & 10 & ON\\ 
3 & 502025010 & 132.1192 & -45.6039 & 2007-07-04 21:41 & 9 & ON\\ 
4 & 502026010 & 132.5157 & -45.5453 & 2007-07-05 02:33 & 10 & ON\\ 
5 & 502027010 & 132.9105 & -45.4880 & 2007-07-05 10:46 & 10 & ON\\ 
6 & 502028010 & 133.3250 & -45.4847 & 2007-07-05 18:01 & 10 & ON\\ 
7 & 502029010 & 133.7759 & -45.5828 & 2007-07-05 23:41 & 13 & ON\\ 
8 & 502030010 & 133.8620 & -45.8612 & 2007-07-06 07:16 & 12 & ON\\ 
9 & 502031010 & 133.4161 & -45.7633 & 2007-07-06 16:50 & 12 & ON\\ 
10 & 502032010 & 132.9993 & -45.7656 & 2007-07-06 22:51 & 12 & ON\\ 
11 & 502033010 & 132.5986 & -45.8258 & 2007-07-07 05:33 & 10 & ON\\ 
12 & 502034010 & 132.2515 & -46.0510 & 2007-07-08 01:43 & 9 & ON\\ 
13 & 502035010 & 131.8510 & -46.1060 & 2007-07-09 03:05 & 8 & ON\\ 
14 & 502036010 & 131.9312 & -46.3861 & 2007-07-09 10:35 & 10 & ON\\ 
15 & 502037010 & 132.3334 & -46.3288 & 2007-07-10 04:40 & 8 & ON\\ 
16 & 502038010 & 132.6844 & -46.1051 & 2007-07-10 11:03 & 15 & ON\\ 
17 & 502039010 & 133.0870 & -46.0455 & 2007-07-10 17:41 & 12 & ON\\ 
18 & 502040010&133.5064 & -46.0415 & 2007-07-10 22:11 & 12 & ON\\ 
19 & 503031010&133.9797 & -46.1476 & 2008-07-03 15:06 & 12 & ON\\ 
20 & 503032010&133.6175 & -46.3272 & 2008-07-04 05:41 & 16 & ON\\ 
21 & 503033010&133.1972 & -46.3295 & 2008-07-04 14:00 & 13 & ON\\ 
22 & 503034010&132.4428 & -46.6127 & 2008-07-05 01:21 & 14 & ON\\
23 & 503035010&132.7881 & -46.3947 & 2008-07-05 09:21 & 15 & ON\\
24 & 503036010&132.0303 & -46.6729 & 2008-07-05 18:31 & 11 & ON\\ 
25 & 503037010&132.5224 & -46.8921 & 2008-07-06 04:01 & 12 & ON\\ 
26 & 503038010&132.8787 & -46.6693 & 2008-07-06 10:25 & 13 & ON\\
27 & 503039010&133.2844 & -46.6061 & 2008-07-06 20:06 & 10 & ON\\ 
28 & 503040010&133.7109 & -46.6037 & 2008-07-07 04:01 & 12 & ON\\ 
29 & 503041010&134.0711 & -46.4300 & 2008-07-07 10:14 & 10 & ON\\
30 & 503042010&134.1665 & -46.7036 & 2008-07-07 17:18 & 9 & ON\\ 
31 & 503043010&133.8003 & -46.8845 & 2008-07-08 02:02 & 10 & ON\\ 
32 & 503044010 & 133.3695 & -46.8865 & 2008-07-08 07:52 & 10 & ON\\
33 & 503045010 & 133.4681 & -47.1616 & 2008-07-08 13:34 & 11 & ON\\ 
34 & 503046010 & 132.1138 & -46.9499 & 2008-07-09 00:13 & 10 & ON\\ 
35 & 503047010 & 132.9660 & -46.9473 & 2008-07-09 06:20 & 11 & ON\\ 
36 & 503048010 & 133.0606 & -47.2239 & 2008-07-09 12:21 & 10 & ON\\ 
37 & 503049010 & 132.6136 & -47.1701 & 2008-07-09 22:01 & 11 & ON\\ 
38 & 503050010 & 132.2003 & -47.2308 & 2008-07-10 05:49 & 13 & ON\\ 
39 & 508036010 & 131.7706 & -46.5645 & 2013-12-04 14:02 & 27 & ON\\ 
40 & 508037010 & 131.7715 & -45.9871 & 2013-11-23 10:38 & 27 & ON\\ 
41 & 508038010 & 131.7730 & -46.2743 & 2013-11-24 01:55 & 33 & ON\\ 
42 & 508060010 & 133.9641 & -46.7818 & 2013-11-25 14:08 & 38 & ON\\ 
43 & 508061010 & 133.8563 & -47.0644 & 2013-11-24 17:37 & 13 & ON\\ 
44 & 508062010 & 133.7724 & -47.2478 & 2013-12-03 23:05 & 26 & ON
\enddata
\tablenotetext{a}{All exposure indicates the effective exposure time of XIS0, XIS1, and XIS3 after processing.}
\end{deluxetable*}

\subsubsection{The ISM Protons}
The interstellar proton column density distribution was derived following the method of \citetalias{2017ApJ...850...71F} by using the NANTEN $^{12}$CO($J = 1$--0) data \citep{2001PASJ...53.1025M} and the \ion{H}{1} data obtained with ATCA \& Parkes \citep{2005ApJS..158..178M}.

These data were smoothed to the same pixels size with the H.E.S.S.\ gamma rays data $\sim 4\farcm 8$.
The velocity integration range was examined and a range from 20\,km\,s$^{-1}$ to 40\,km\,s$^{-1}$ was adopted in the present work as explained below in this subsection.
The CO-to-H$_{2}$ conversion factor $X_{\mathrm{CO}}$ was taken to be $1.5\times 10^{20}$\,cm$^{-2}$\,(K\,km\,s$^{-1}$)$^{-1}$ \citep{2022ApJ...938...94A}.
The \ion{H}{1} column density $N(\mbox{\ion{H}{1}})$ was calculated by considering the \ion{H}{1} optical depth effect based on the sub-mm dust optical depth following \citetalias{2017ApJ...850...71F}.
The total ISM proton column density $N_{\mathrm{p}}$ is given as follows;
\begin{equation}
N_{\mathrm{p}} = N(\mbox{\ion{H}{1}}) + 2N(\mathrm{H}_{2})
\end{equation}
and re-gridded in a pixel size of $11\farcm 4$.
Here $N$(H$_2$) is the column density of molecular hydrogen H$_2$.

It is generally a complicated task to identify the ISM that is associated with an SNR.
In case of RXJ0852, it was obvious that the CO and \ion{H}{1} gas at velocity around 25--30\,km\,s$^{-1}$ are associated with the SNR \citepalias{2017ApJ...850...71F}.
The gas in this velocity range shows good morphological correspondence with the X-ray shell in the south-western part \citepalias[see Figures 1(b)-3 of][]{2017ApJ...850...71F}.
The velocity range is in conflict with the gas obeying the galactic rotation at a distance near 1\,kpc.
The gas motion is significantly affected by the local disturbance in the order of $>10$\,km\,s$^{-1}$, which is likely driven by the supershells in the region of RXJ0852 as identified in \citetalias{2017ApJ...850...71F}.
In the present work, we took a velocity range of a velocity width of 20\,km\,s$^{-1}$, i.e., 20--40\,km\,s$^{-1}$, which corresponds to the expanding pattern of the shell created by a stellar-wind bubble as shown in Appendix \ref{sec:associated_ISM}.
 The velocity span is the same with that of RXJ1713 \citepalias{2012ApJ...746...82F,2021ApJ...915...84F}, and is consistent with that expected if it is created by the stellar-wind acceleration.
\citetalias{2017ApJ...850...71F} suggested that gas with an even larger velocity up to 50\,km\,s$^{-1}$ might contribute to the gamma ray emission, while the increase in \ion{H}{1} mass for the larger velocity range is unimportant, less than $\sim 10$\% of the total \ion{H}{1} mass.

\subsection{RXJ1713}\label{rxj1713:data}
We used the data of the gamma rays, X rays and the interstellar protons $N_{\mathrm{p}}$ which were published in \citetalias{2012ApJ...746...82F} and \citetalias{2021ApJ...915...84F}.
The gamma rays are obtained by H.E.S.S.\ as shown by the excess count map at an energy range $E > 2$\,TeV \citep{2018A&A...612A...6H} which is binned at FWHM resolution of $4\farcm 8$ ($\sim 1.4$\,pc).
The X rays are obtained by XMM-Newton in an energy range of 1--5\,keV \citep{2004ASPC..314..759G}, where the two compact sources 1WGA~J1714.4$-$3945 and 1WGA~J1713.4$-$394 are masked, and binned at $4\farcm 8$ grid \citep{2021ApJ...915...84F}.
The ISM proton column density $N_{\mathrm{p}}$ consists of the NANTEN $^{12}$CO($J = 1$--0) \citep{2005ApJ...631..947M} and the ATCA \& Parkes \ion{H}{1} \citep{2005ApJS..158..178M}, and is also binned at $4\farcm 8$ grid.

\begin{figure*}[p]
\begin{center}
\includegraphics[width=\linewidth,clip]{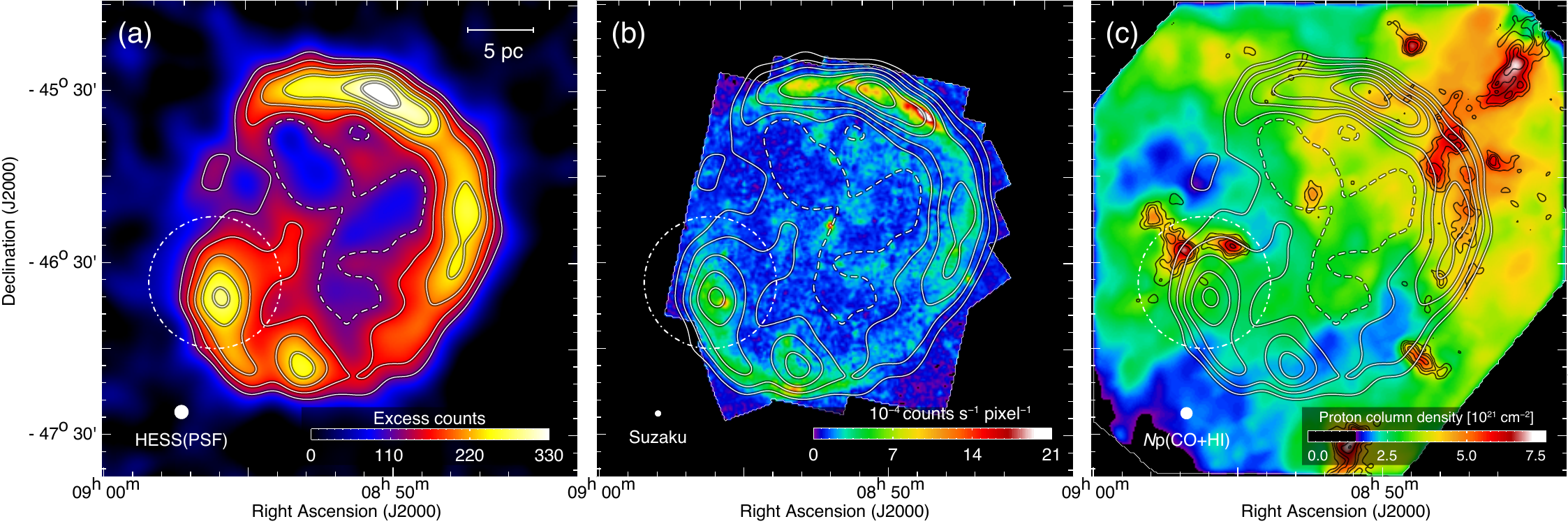}
\end{center}
\caption{
Maps of (a) TeV gamma ray excess counts ($E>100$\,GeV) \citep{2018A&A...612A...7H}, (b) X ray counts ($E=2.0$--5.7\,keV) \citepalias{2017ApJ...850...71F}, and (c) ISM proton column density (this work) toward RXJ0852.
The superposed white contours in Panels (a)--(c) indicate excess counts of TeV gamma rays (the lowest level and intervals are 120 and 40 counts, respectively). 
The black contours in Panel (c) indicate $^{12}$CO($J=1\mbox{--}0$) intensity integrated in a velocity range from 20 to 40\,km\,s$^{-1}$ (the contour levels are 2.5, 5.0, 7.5, 12.5, and 17.5\,K\,km\,s$^{-1}$).
The dash-dotted circle outlines the PWN unrelated to the SNR \citep[see][]{2018A&A...612A...7H}.
}
\label{fig:Gammaray_Xray_ISMNp_img_Gammaraycont}
\end{figure*}

\begin{figure*}
\begin{center}
\includegraphics[width=\linewidth,clip]{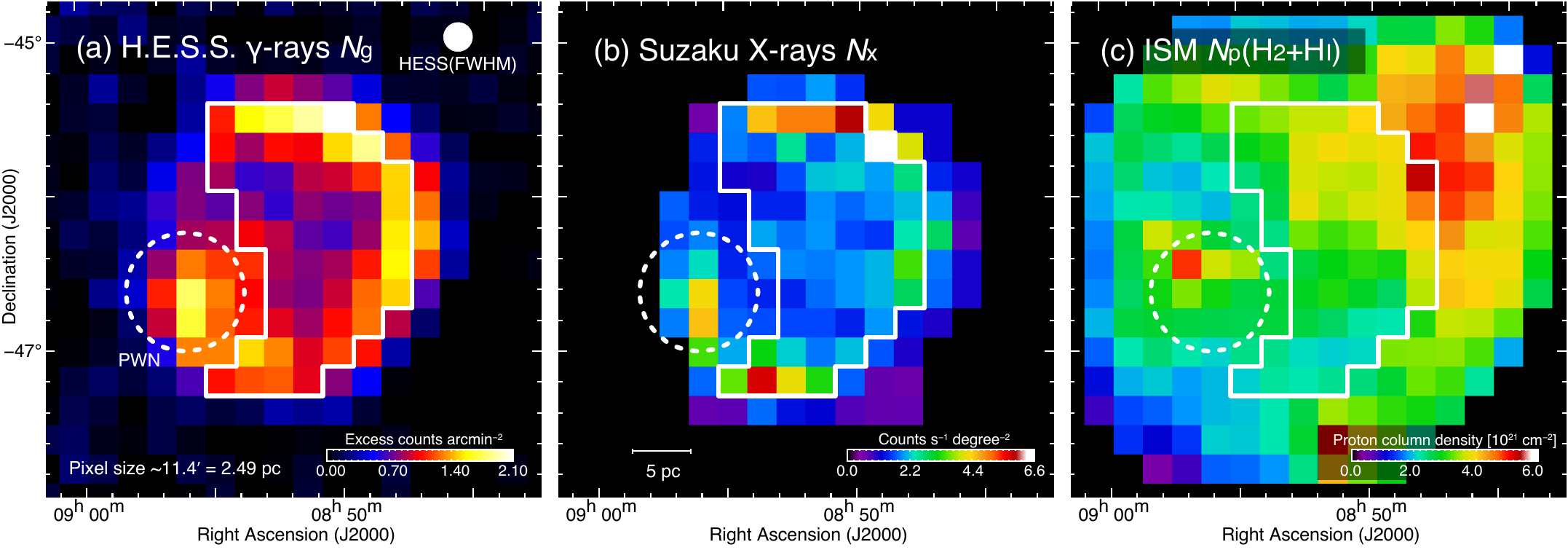}
\end{center}
\caption{Spatial distribution of (a) $N_{\mathrm{g}}$, (b) $N_{\mathrm{x}}$, and (c) $N_{\mathrm{p}}$ for RXJ0852.
The three datasets are pixelated on an $11\farcm 4$ grid to match the H.E.S.S. resolution.
The white polygon in each panel indicates the region of interest for the present analysis, and the dashed circle outlines the PWN unrelated to the SNR.
}
\label{fig:ng_nx_np_0852}
\end{figure*}

\begin{figure*}[p]
\gridline{\fig{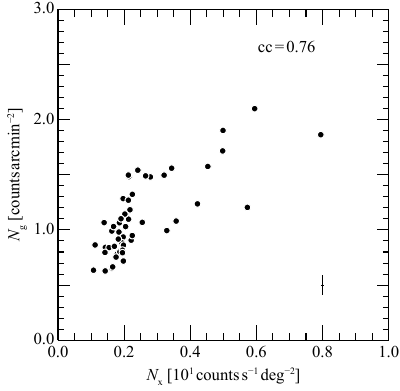}{0.32\textwidth}{(a)}
          \fig{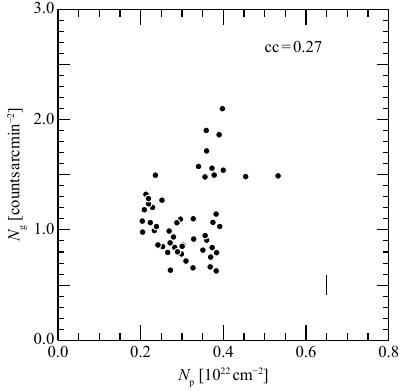}{0.32\textwidth}{(b)}
          \fig{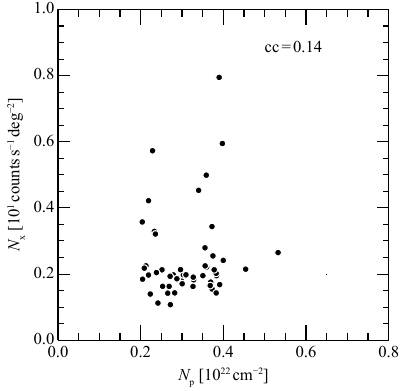}{0.32\textwidth}{(c)}}
\caption{
Scatter plots of (a) $N_{\mathrm{g}}$-$N_{\mathrm{x}}$, (b) $N_{\mathrm{g}}$-$N_{\mathrm{p}}$, and (c) $N_{\mathrm{p}}$-$N_{\mathrm{x}}$ for RXJ0852.
The correlation coefficient is noted in the upper right corner of each panel, and the horizontal/vertical error bars in the bottom-right corner present the median values of the uncertainties.}
\label{fig:NgNxNp_all_plot_error}
\end{figure*}%

\begin{figure*}
\begin{center}
\includegraphics[width=\linewidth]{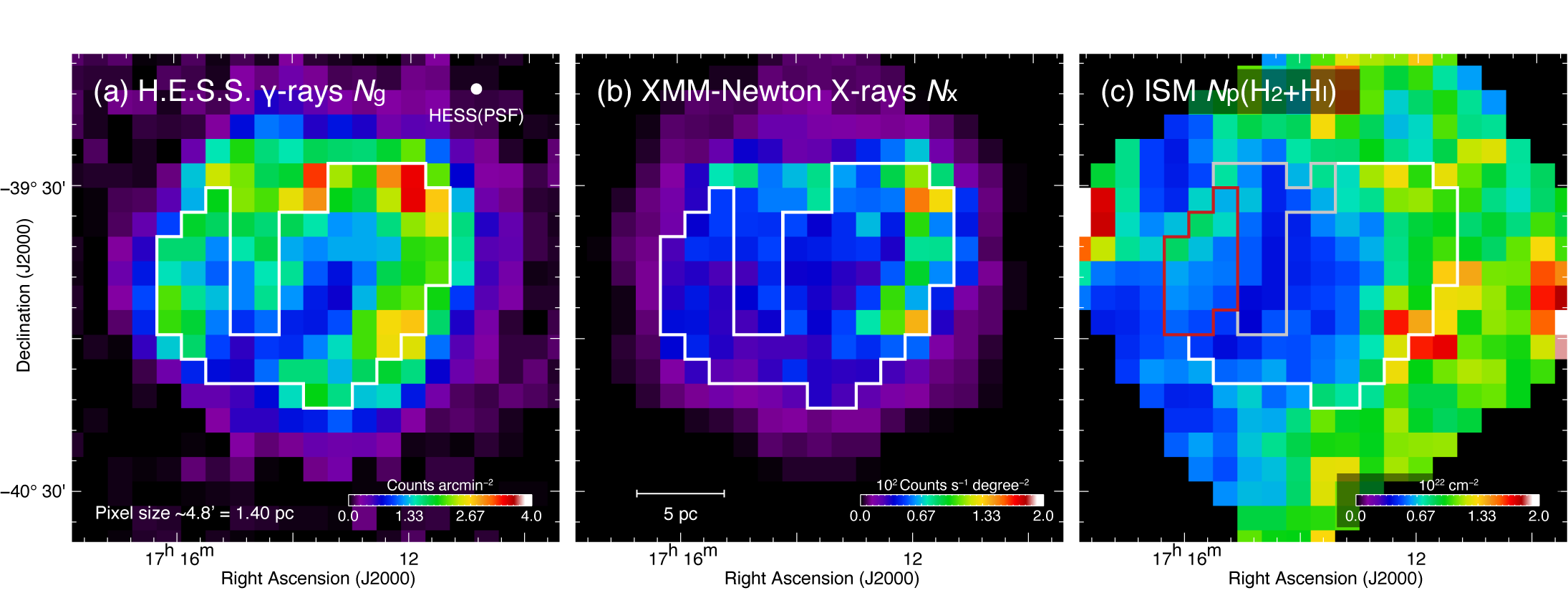}
\end{center}
\caption{Spatial distribution of (a) $N_{\mathrm{g}}$, (b) $N_{\mathrm{x}}$, and (c) $N_{\mathrm{p}}$ for RXJ1713 \citepalias[identical to the H.E.S.S.18 ($E>2$\,TeV) $4\farcm 8$ resolution dataset in][]{2021ApJ...915...84F}.
The white polygon in each panel indicates the region of interest for the present analysis.
The red polygon in Panel (c) outlines the region where \ion{H}{1} absorption is corrected \citepalias[see Section 2.3 of][]{2021ApJ...915...84F}.
}\label{fig:RXJ1713_NgNxNp_map}
\end{figure*}

\begin{figure*}
\gridline{\fig{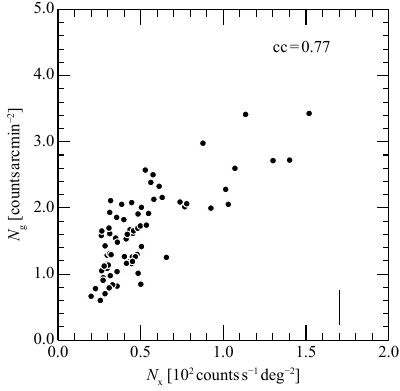}{0.32\textwidth}{(a)}
          \fig{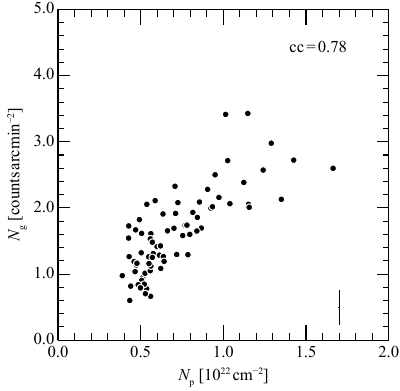}{0.32\textwidth}{(b)}
          \fig{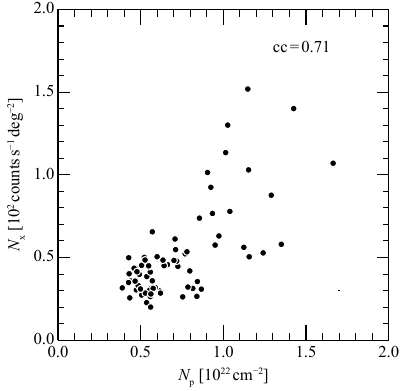}{0.32\textwidth}{(c)}}
\caption{
Same as Figure \ref{fig:NgNxNp_all_plot_error} but for RXJ1713.}
\label{fig:RXJ1713_NgNxNp_all_plot_error}
\end{figure*}%

\subsection{The Distributions of the Three Quantities}
\subsubsection{RXJ0852}
The three physical quantities relevant to the analysis including the H.E.S.S.\ TeV gamma ray count $N_{\mathrm{g}}$, the \textit{Suzaku} nonthermal X ray counts $N_{\mathrm{x}}$, and the interstellar proton column density $N_{\mathrm{p}}$.
Figure \ref{fig:Gammaray_Xray_ISMNp_img_Gammaraycont}(a) shows the distribution of gamma-rays.
Figures \ref{fig:Gammaray_Xray_ISMNp_img_Gammaraycont}(b) and \ref{fig:Gammaray_Xray_ISMNp_img_Gammaraycont}(c) show the distributions of the nonthermal X-rays and the ISM protons superposed with the TeV gamma-ray contours.
Figure \ref{fig:ng_nx_np_0852} shows the three datasets gridded to a pixel size of $11\farcm 4$, which is matched to the H.E.S.S.\ resolution as given in Table \ref{tab:summary_multireg}.
The pixels which are analyzed are enclosed by white solid lines.
We excluded the pixels toward the PWN unrelated to the SNR in the east (outlined by the dashed-line circle in Figure \ref{fig:ng_nx_np_0852}).
Figure \ref{fig:NgNxNp_all_plot_error} shows three scatter plots of $N_{\mathrm{g}}$-$N_{\mathrm{x}}$, $N_{\mathrm{g}}$-$N_{\mathrm{p}}$, and $N_{\mathrm{p}}$-$N_{\mathrm{x}}$.
Figure \ref{fig:NgNxNp_all_plot_error}(a) shows that $N_{\mathrm{g}}$ has a high correlation with $N_{\mathrm{x}}$ with a correlation coefficient of 0.76.
On the other hand, Figure \ref{fig:NgNxNp_all_plot_error}(b) shows that $N_{\mathrm{g}}$-$N_{\mathrm{p}}$ has a lower correlation coefficient of 0.27 with a larger scatter than in $N_{\mathrm{g}}$-$N_{\mathrm{x}}$.
In Figure \ref{fig:NgNxNp_all_plot_error}(c) $N_{\mathrm{p}}$-$N_{\mathrm{x}}$ has a positive correlation coefficient of 0.14.

\subsubsection{RXJ1713}
Figure \ref{fig:RXJ1713_NgNxNp_map} shows the distributions of the three physical quantities relevant in the analysis include the H.E.S.S.\ TeV gamma ray count $N_{\mathrm{g}}$, the XMM-Newton nonthermal X ray counts $N_{\mathrm{x}}$, and the interstellar proton column density $N_{\mathrm{p}}$.
Figures \ref{fig:RXJ1713_NgNxNp_map}(a)--(c) show the pixels of $N_{\mathrm{g}}$, $N_{\mathrm{x}}$ and $N_{\mathrm{p}}$, respectively.
The analyzed pixels are enclosed by white solid lines.
Figures \ref{fig:RXJ1713_NgNxNp_all_plot_error}(a)--(c) show three scatter plots of $N_{\mathrm{g}}$-$N_{\mathrm{x}}$, $N_{\mathrm{g}}$-$N_{\mathrm{p}}$ and $N_{\mathrm{p}}$-$N_{\mathrm{x}}$.
Figure \ref{fig:RXJ1713_NgNxNp_all_plot_error}(a) shows that $N_{\mathrm{g}}$ has a high correlation with $N_{\mathrm{x}}$ with a correlation coefficient of 0.77, and Figure \ref{fig:RXJ1713_NgNxNp_all_plot_error}(b) also shows that $N_{\mathrm{g}}$-$N_{\mathrm{p}}$ has a high correlation coefficient of 0.78.
The increase of $N_{\mathrm{g}}$ with both of $N_{\mathrm{x}}$ and $N_{\mathrm{p}}$ suggests that $N_{\mathrm{g}}$ increases with both of $N_{\mathrm{p}}$ and $N_{\mathrm{x}}$.
Figure \ref{fig:RXJ1713_NgNxNp_all_plot_error}(c) shows that $N_{\mathrm{p}}$-$N_{\mathrm{x}}$ has a positive correlation coefficient of 0.71.

\section{Formulation and Fitting}\label{sec:formulation_and_fitting}
\subsection{Formulation}\label{subsec:formulation_and_fitting:formulation}

Following \citetalias{2021ApJ...915...84F}, we express $N_\mathrm{g}$ by a linear combination of $N_\mathrm{x}$ and $N_\mathrm{p}$ as follows;
\begin{eqnarray}\label{eqn:regression_model}
N_{\mathrm g} = a N_{\mathrm p} + b N_{\mathrm x}.
\end{eqnarray}
The first term stands for the hadronic gamma rays via the p-p reaction, $N_{\mathrm{g}}^{\mathrm{hadronic}}$, and the second term for the leptonic gamma rays, $N_{\mathrm{g}}^{\mathrm{leptonic}}$, via the IC scattering.
We assume here that the energy density of CR protons is uniform within the volume, where the energy density of the CR electrons is negligibly small due to a small electron/proton ratio or with some non-uniformity (\citetalias[cf.,][]{2010ApJ...708..965Z};\citealt{2021A&A...654A.139B}).
In addition, we assume that the magnetic field strength $B$ is uniform within the volume as suggested by the numerical simulations by \citet{2012ApJ...744...71I}, while the simulations show a moderate variation of the field strength by $\sim 20$\% when averaged in volume.
The CMB photon density is uniform and additional stellar photons are negligible in the region of RXJ0852 as shown by no \ion{H}{2} regions or OB stars around the region \citep{2001PASJ...53.1025M}.

In the following we give approximate relationships of the quantities in Equation (\ref{eqn:regression_model}).
The first term is expressed by the CR proton volume density $n_{\mathrm{p}}(\mathrm{CR})$ times the ISM proton column density $N_{\mathrm{p}}$,
\begin{eqnarray}
N_{\mathrm{g}}^{\mathrm{hadronic}} = k_{1} n_{\mathrm p}(\mathrm{CR}) N_{\mathrm p} = a N_{\mathrm p},
\end{eqnarray}
where $a$ is a constant including the p-p reaction coefficient ($k_{1}$) times $n_{\mathrm{p}}(\mathrm{CR})$, and $k_{1}$ includes a branching ratio of 1/3 for the $\pi^{0}$ production.
The second term is proportional to the low-energy photon volume density $n(\mathrm{CMB})$ times the CR electron column density $N_{\mathrm{e}}(\mathrm{CR})$,
\begin{eqnarray}
N_{\mathrm{g}}^{\mathrm{leptonic}} = k_{2} N_{\mathrm e}(\mathrm {CR}) n(\mathrm {CMB}) = b N_{\mathrm x}, 
\end{eqnarray}
and $N_{\mathrm{x}}$ is expressed by the CR electron column density $N_{\mathrm{e}}(\mathrm{CR})$ and the $B$ field as follows,
\begin{eqnarray}
N_{\mathrm x} = k_{3} N_{\mathrm e}(\mathrm {CR}) B^{2}. 
\end{eqnarray}
Here, $b = k_{2} n(\mathrm{CMB})/[k_{3} B^{2}]$ is a constant which consists of the IC scattering coefficient $k_2$ times $n(\mbox{CMB}){k_{3}}^{-1}B^{-2}$, $k_{3}$ is the synchrotron emissivity coefficient, and $N_{\mathrm{e}}(\mbox{CR})$ is a sum of the cosmic ray electron volume density $n_{\mathrm{e}}(\mbox{CR})$ along the line of sight.

\begin{figure*}[p]
\gridline{\fig{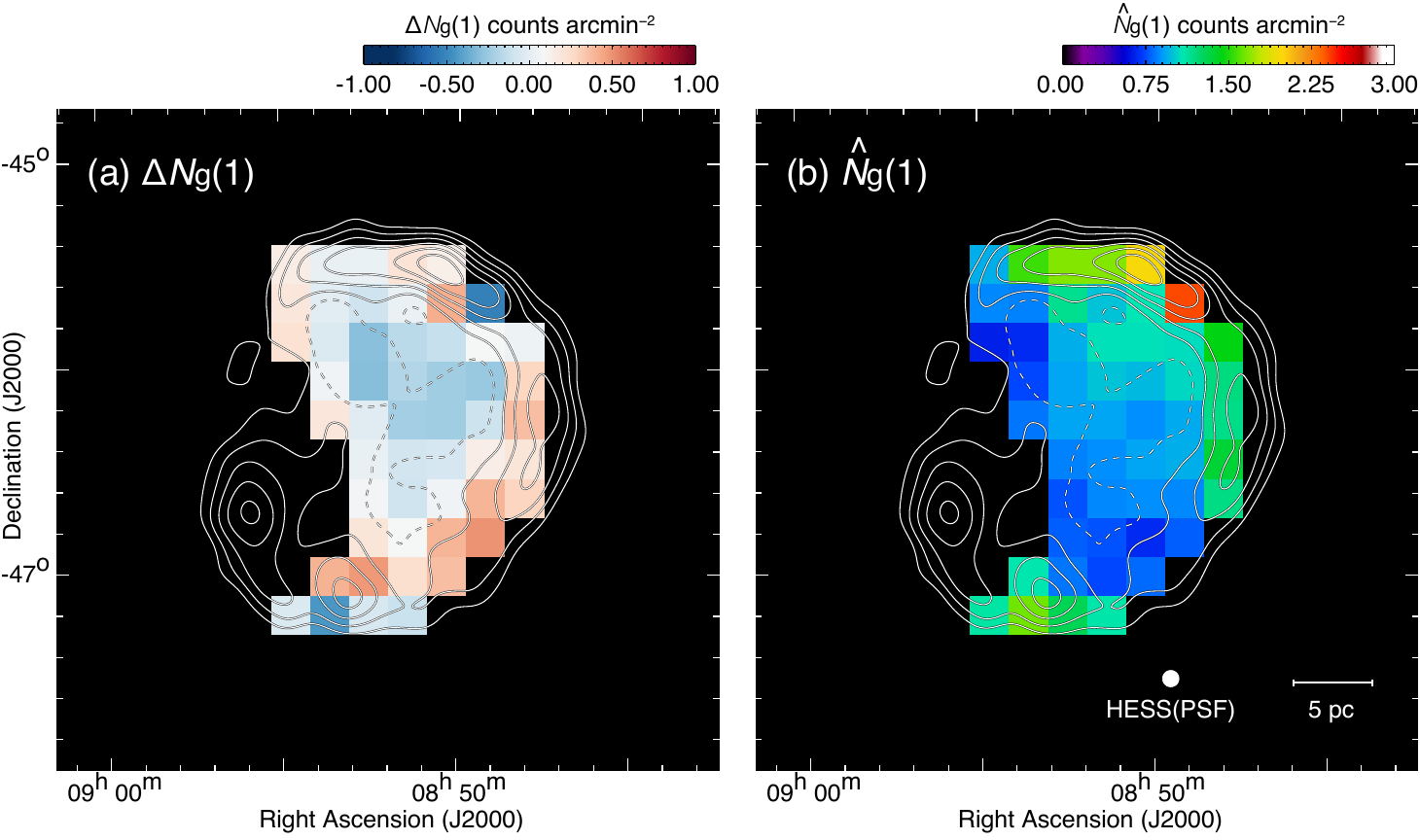}{0.8\textwidth}{}}
\gridline{\fig{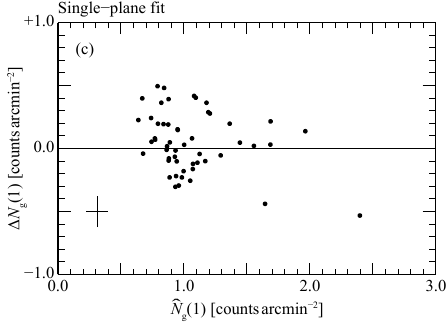}{0.5\textwidth}{}}
\caption{Results of the single-plane fit for RXJ0852.
(a) Maps of $\Delta N_{\mathrm{g}}(1)$ derived from Equation (\ref{eqn:deltaNg}) and (b) $\widehat{N}_{\mathrm{g}}(1)$ derived from Equation (\ref{eqn:regression_model}).
The superposed contours in both Panels (a) and (b) indicate excess counts of TeV gamma rays (identical to those in Figure \ref{fig:Gammaray_Xray_ISMNp_img_Gammaraycont}).
(c) A plot of $\Delta N_{\mathrm{g}}(1)$ with respect to $\widehat{N}_{\mathrm{g}}(1)$.
The horizontal- and vertical error bars shown in the left bottom corner indicate the median values of $\sigma(\widehat{N}_{\mathrm{g}}(1))$ and $\sigma(\Delta N_{\mathrm{g}}(1))$ \citepalias[see Equations (6) and (7) in][]{2021ApJ...915...84F}, respectively.
}
\label{fig:dNg_Nghat_all_plot_error_map}
\end{figure*}%

\begin{figure*}
\gridline{\fig{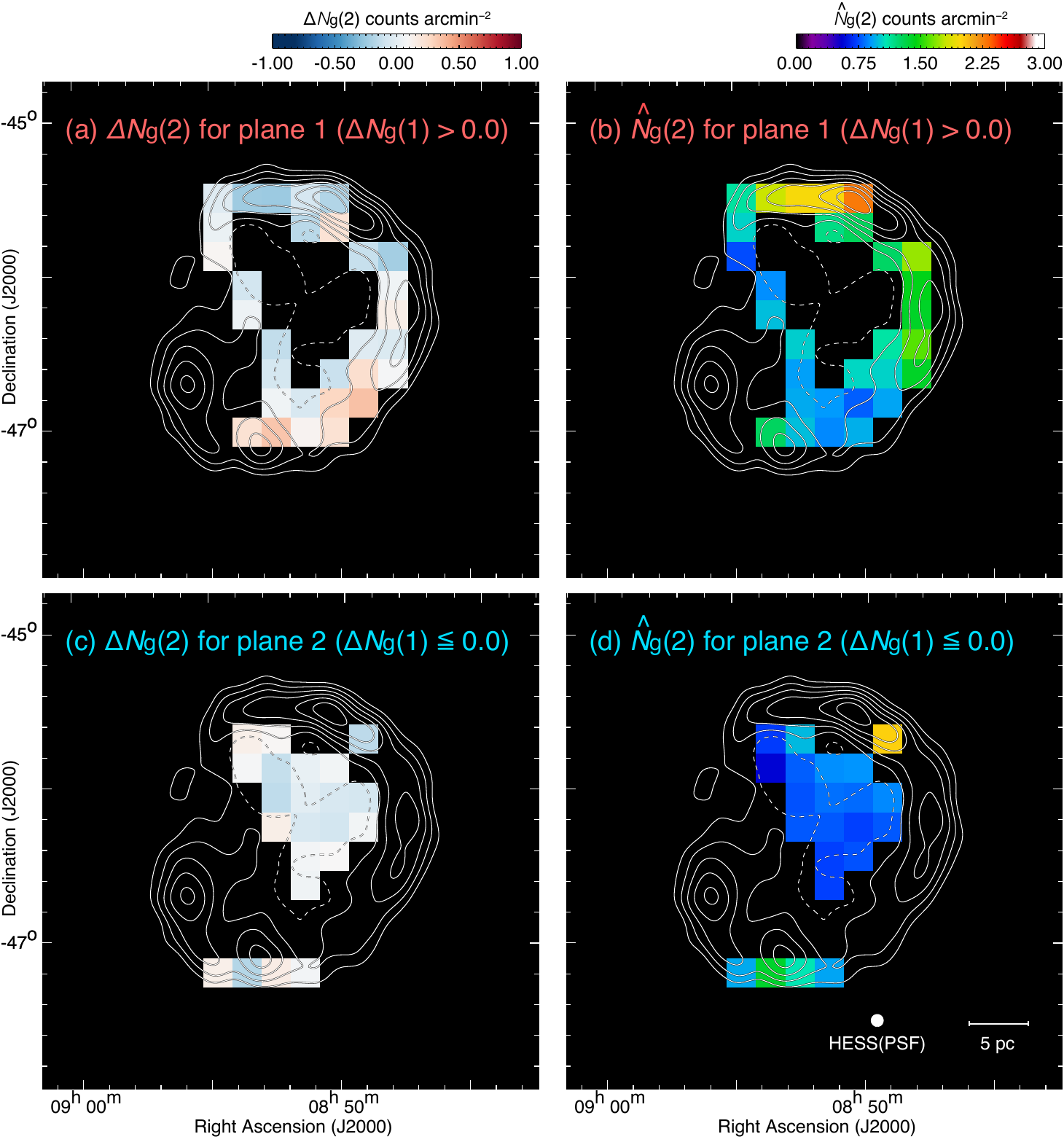}{0.7\textwidth}{}}
\gridline{\fig{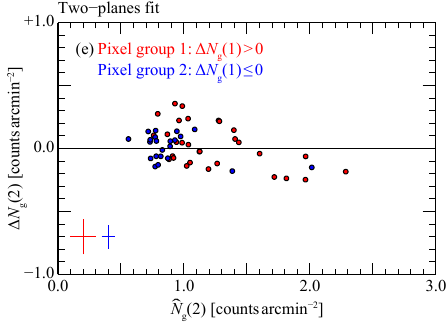}{0.5\textwidth}{}}
\caption{Results of the two-planes fit for RXJ0852.
(a)--(d) Maps of $\Delta N_{\mathrm{g}}(2)$ (left panels) and $\widehat{N}_{\mathrm{g}}(2)$ (right panels) for the pixel groups 1 ($\Delta N_{\mathrm{g}}(1)>0$, top panels) and 2 ($\Delta N_{\mathrm{g}}(1)\le 0$, middle panels).
The superposed contours in each panel are identical to those in Figure \ref{fig:Gammaray_Xray_ISMNp_img_Gammaraycont}.
(e) A plot of $\Delta N_{\mathrm{g}}(2)$ with respect to $\widehat{N}_{\mathrm{g}}(2)$.
The horizontal- and vertical error bars shown in the left bottom corner indicate the median values of $\sigma(\widehat{N}_{\mathrm{g}}(2))$ and $\sigma(\Delta N_{\mathrm{g}}(2))$, respectively.
The red circles and error bars are for the pixel group 1 and blue ones are for the pixel group 2.
}
\label{fig:dNg_all_inner_shell_plot_error_map}
\end{figure*}%

\begin{figure*}
\gridline{\fig{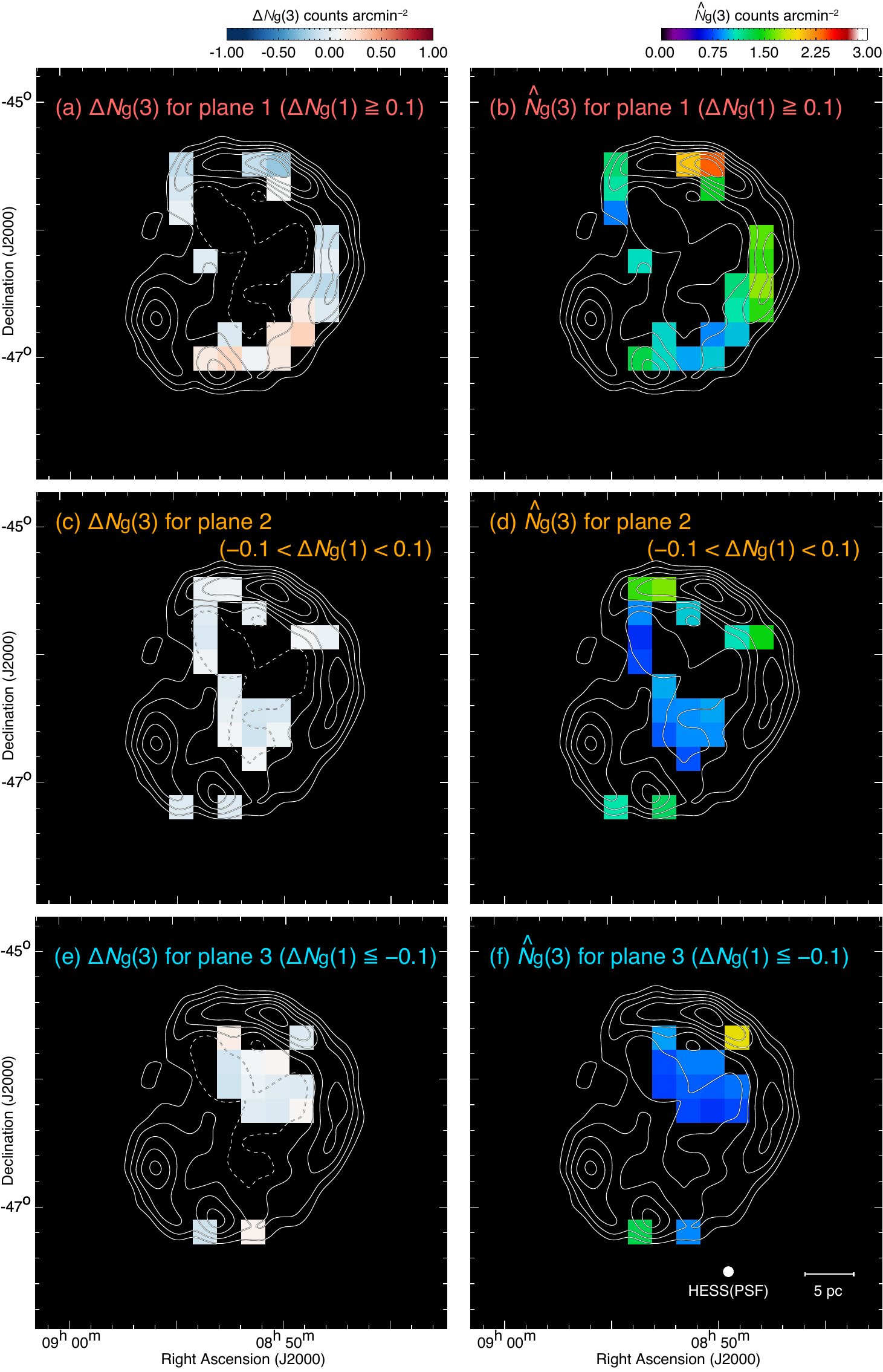}{.8\textwidth}{}}
\caption{Results of three-planes fit for RXJ0852.
(a)--(f) Maps of $\Delta N_{\mathrm{g}}(3)$ (left panels) and $\widehat{N}_{\mathrm{g}}(3)$ (right panels) for the pixel groups 1 ($\Delta N_{\mathrm{g}}(1)> 0.1$, top panels), 2 ($-0.1 \le \Delta N_{\mathrm{g}}(1) \le 0.1$, second row panels) and 3 ($\Delta N_{\mathrm{g}}(1) < -0.1$, third row panels).
The superposed contours in each panel are identical to those in Figure \ref{fig:Gammaray_Xray_ISMNp_img_Gammaraycont}.
}
\label{fig:dNg_plot_all_3region_dNgpm0p1_error_map}
\end{figure*}

\addtocounter{figure}{-1}
\begin{figure*}
\gridline{\fig{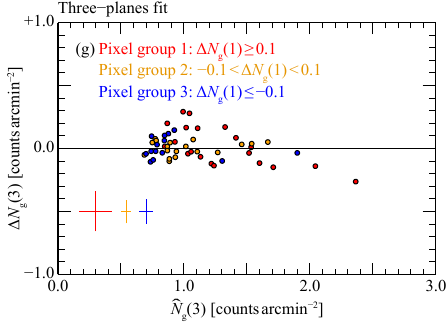}{0.5\textwidth}{}}
\caption{Continued.
(g) A plot of $\Delta N_\mathrm{g}(3)$ with respect to $\widehat{N}_{\mathrm{g}}(3)$.
The horizontal- and vertical error bars shown in the left bottom corner indicate the median values of $\sigma(\widehat{N}_{\mathrm{g}}(3))$} and $\sigma(\Delta N_{\mathrm{g}}(3))$, respectively.
The red circles and error bars are for the pixel group 1, the orange and blue ones are for the pixel groups 2 and 3, respectively.
\end{figure*}

\begin{figure}
\gridline{\fig{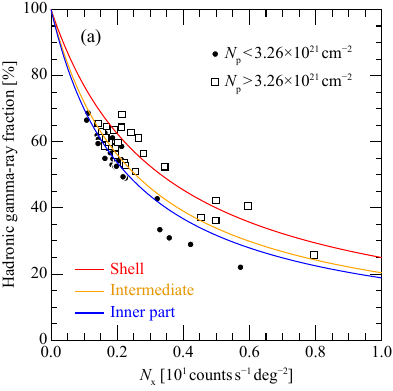}{0.9\linewidth}{}}
\gridline{\fig{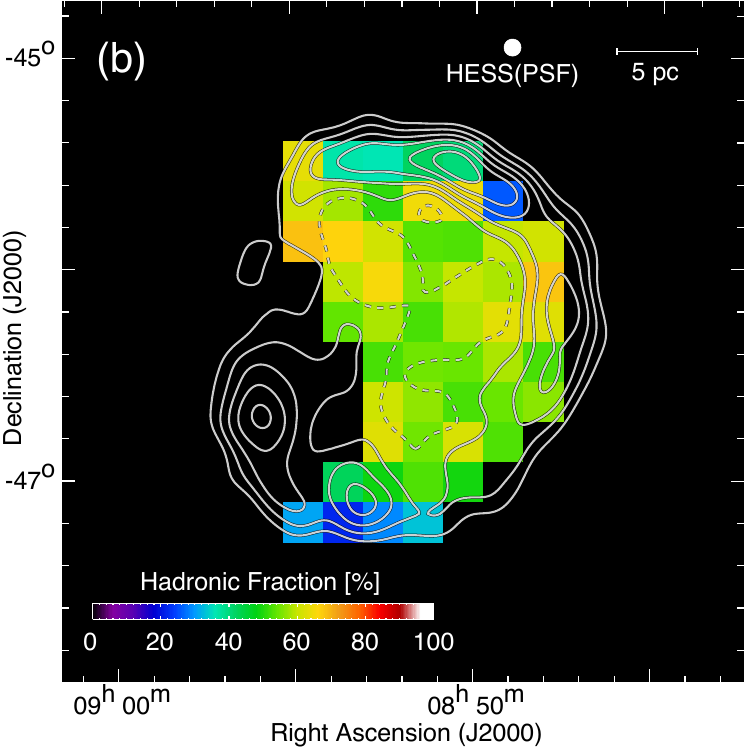}{0.9\linewidth}{}}
\caption{(a) Scatter plot between hadronic gamma-ray fraction and $N_{\mathrm{x}}$ for RXJ0852.
The filled circles and open squares represent the data points where $N_{\mathrm{p}}<3.26\times 10^{21}$\,cm$^{-2}$ and $<3.26\times 10^{21}$\,cm$^{-2}$, respectively.
The overlaid solid lines show the relationships derived by an equation $aN_{\mathrm{p}}/(aN_{\mathrm{p}}+bN_{\mathrm{x}})$ for $N_{\mathrm{p}} = 3.26\times 10^{21}$\,cm$^{-2}$.
The red, orange, and blue lines are for the shell, intermediate, and inner parts of the SNR, respectively (the $a$ and $b$ values for each part are in Table \ref{tab:regression_results_0852}).
(b) Distributions of hadronic gamma-ray fraction overlaid with gamma-ray contours (identical to those in Figure \ref{fig:Gammaray_Xray_ISMNp_img_Gammaraycont}).}
\label{fig:hadronicgammafraction_all_kttv_div5_plotmap}
\end{figure}

\begin{deluxetable*}{lRRRRR}
\tablecaption{Summary of the Multiple Linear Regression for RXJ0852} \label{tab:regression_results_0852}
\tablewidth{0pt}
\tablehead{
\colhead{Model} & \colhead{$n$} & \colhead{$a$} & \colhead{$b$} & \colhead{$\chi ^{2} / \nu $} & \colhead{VIF}
}
\decimalcolnumbers
\startdata
Single-plane fit & 52 & 1.60\pm 0.21 & 22.3\pm 2.64 & 6.65 & 1.0 \\
\hline
\multicolumn{5}{l}{Two-plane fit} \\
\hline
Shell ($\Delta N_{\mathrm{g}}(1)>0$) & 30 & 1.97\pm 0.26 & 25.2\pm 3.29 & 3.13 & 1.2 \\
Inner part ($\Delta N_{\mathrm{g}}(1)\le 0$) & 22 & 1.31\pm 0.12 & 18.9\pm 1.58 & 1.64 & 1.0 \\
\hline
\multicolumn{5}{l}{Three-plane fit} \\
\hline
Shell ($\Delta N_{\mathrm{g}}(1) \ge 0.1$) & 20 & 2.41\pm 0.33 & 23.6\pm 3.93 & 2.44 & 1.3 \\
Intermediate ($-0.1 < \Delta N_{\mathrm{g}}(1) < 0.1$) & 18 & 1.68\pm 0.10 & 21.4\pm 1.32 & 0.47 & 1.0 \\
Inner part ($\Delta N_{\mathrm{g}}(1) \le -0.1$) & 14 & 1.26\pm 0.12 & 17.7\pm 1.51 & 1.15 & 1.3 \\
\enddata
\tablecomments{
Columns (1): model name; (2): number of pixels of the data set; (3) and (4): estimated regression coefficients and their standard deviations; (5): reduced-$\chi^{2}$ (Equation (\ref{eqn:reduced-chi-square})); (6): variance inflation factor, less than a severe threshold of 3 indicates that multicollinearity is not a significant issue in the analysis \citepalias[see Appendix C3 of][]{2021ApJ...915...84F}.
}
\end{deluxetable*}

\begin{deluxetable*}{lRRRRRRR}
\tablecaption{The hadronic and leptonic gamma-ray components in RXJ0852} \label{tab:estimate_hadron_and_lepton_0852}
\tablewidth{0pt}
\tablehead{ & & & \multicolumn2c{Hadronic component} & &  \multicolumn2c{Leptonic component} \\ 
\cline{4-5}
\cline{7-8}
\colhead{Model} & \colhead{$\langle\widehat{N}_{\mathrm{g}}\rangle$} & \colhead{$\langle{N}_{\mathrm{p}}\rangle$} & \colhead{$\langle \widehat{N}_{\mathrm{g}}^{\mathrm{hadronic}}\rangle$} & \colhead{$\langle\widehat{N}_{\mathrm{g}}^{\mathrm{hadronic}}\rangle/\langle\widehat{N}_{\mathrm{g}}\rangle$} & \colhead{$\langle N_{\mathrm{x}}\rangle$} & \colhead{$\langle\widehat{N}_{\mathrm{g}}^{\mathrm{leptonic}}\rangle$} & \colhead{$\langle\widehat{N}_{\mathrm{g}}^{\mathrm{leptonic}}\rangle/\langle\widehat{N}_{\mathrm{g}}\rangle$} }
\decimalcolnumbers
\startdata
Single-plane fit & 1.06\pm 0.01 & 0.31 & 0.50\pm 0.01 & (47 \pm 1)\% &  0.25 & 0.56\pm 0.01 & (53\pm 1)\%  \\
\hline
\multicolumn{8}{l}{Two-plane fit} \\
\hline
Whole & 1.09\pm 0.01 & 0.31 & 0.53\pm 0.01 & (48\pm 1)\% & 0.25 & 0.56\pm 0.01 & (52\pm 1)\%  \\
Shell ($\Delta N_{\mathrm{g}}(1)>0$) & 1.23\pm 0.02 & 0.31 & 0.61\pm 0.02 & (50\pm 2)\% & 0.25 & 0.62\pm 0.02 & (50\pm 2)\%  \\
Inner part ($\Delta N_{\mathrm{g}}(1)\le 0$) & 0.90\pm 0.01 & 0.32 & 0.42\pm 0.01 & (46\pm 1)\% & 0.25 & 0.48\pm 0.01 & (54\pm 1)\% \\
\hline
\multicolumn{8}{l}{Three-plane fit} \\
\hline
Whole & 1.09\pm 0.01 & 0.31 & 0.57\pm 0.01 & (52\pm 1)\% & 0.25 & 0.53\pm 0.01 & (48\pm 1)\%  \\
Shell ($\Delta N_{\mathrm{g}}(1) \ge 0.1$) & 1.30\pm 0.03 & 0.30 & 0.71\pm 0.02 & (55\pm 2)\% & 0.25 & 0.59\pm 0.02 & (45\pm 2)\% \\
Intermediate ($|\Delta N_{\mathrm{g}}(1)| < 0.1$) & 1.02\pm 0.01 & 0.31 & 0.52\pm 0.01 & (51\pm 1)\% & 0.23 & 0.50\pm 0.01 & (49\pm 1)\% \\
Inner part ($\Delta N_{\mathrm{g}}(1) \le -0.1$) & 0.91\pm 0.02 & 0.34 & 0.43\pm 0.01 & (47\pm 2)\% & 0.27 & 0.48\pm 0.01 & (53\pm 2)\% \\
\enddata
\tablecomments{
Columns (1): model name; (2), (3) and (6): spatial averages of observed $N_{\mathrm{g}}$ (counts\,arcmin$^{-2}$), $N_{\mathrm{p}}$ ($10^{22}$\,cm$^{-2}$) and $N_\mathrm{x}$ ($10^{1}$\,counts\,s$^{-1}$\,degree$^{-2}$); (4): spatial average of the predicted hadronic-origin gamma-rays (counts\,arcmin$^{-2}$) given by $\sum_{i}^{n}(\widehat{N}_{\mathrm{g}, i}^{\mathrm{hadronic}})/n$ and its standard deviation; (5): fraction of the hadronic component; (7) and (8): same as (4) and (5) but for the predicted leptonic-origin gamma-rays.
}
\end{deluxetable*}

\subsection{Fitting of the Hadronic and Leptonic Components in RXJ0852}
In order to determine the best fit flat plane(s) in the 3D space of $N_{\mathrm{p}}$-$N_{\mathrm{x}}$-$N_{\mathrm{g}}$, we applied Equation (\ref{eqn:regression_model}) to all the data pixels in a 3D space by the least-squares fitting following \citetalias{2021ApJ...915...84F}.

We obtained a multiple linear regression plane as summarized in the first line in Table \ref{tab:regression_results_0852}, a single plane as adopted in \citetalias{2021ApJ...915...84F}.
The distribution of the difference in $N_{\mathrm{g}}$ is defined as
\begin{equation}\label{eqn:deltaNg}
    \Delta N_{\mathrm{g}}(1) = N_{\mathrm{g}}-\widehat{N}_{\mathrm{g}}(1),
\end{equation}
where the 1 in the brackets stands for the single-plane and the hat-symbol ($\widehat{\phantom{N}}$) for the predicted value by the fit plane.
The spatial distribution of $\Delta N_{\mathrm{g}}(1)$ and $\widehat{N}_{\mathrm{g}}(1)$ are shown in Figures \ref{fig:dNg_Nghat_all_plot_error_map}(a) and (b), respectively.
We see that $\Delta N_{\mathrm{g}}(1)$ tends to become large in the shell especially in the south, and to become small in the inner part.
Figure \ref{fig:dNg_Nghat_all_plot_error_map}(c) shows a plot of $\Delta N_{\mathrm{g}}(1)$ as a function of $\widehat{N}_{\mathrm{g}}$.
Table \ref{tab:regression_results_0852} shows the fit results of $a$ and $b$ as well as the reduced-$\chi^{2}$ in the fit
\begin{equation}\label{eqn:reduced-chi-square}
    \frac{\chi^{2}}{\nu}=\frac{1}{\nu}\sum_{i=1}^{n}\frac{{\Delta N_{\mathrm{g}, i}}^{2}}{\sigma(N_{\mathrm{g}, i})^{2}},
\end{equation}
where, $\nu=n-2$ is the degree of freedom, $n$ is the number of pixels of the data set, and $\sigma(N_{\mathrm{g}, i})$ is the uncertainty of $N_{\mathrm{g}, i}$ estimated from the statistical Poisson error in pixel counts.
The subscript $i$ stands for the $i$-th pixel of the data set.
We find that the reduced-$\chi^{2}$ is significantly large, more than 6, which is unacceptable.
This is probably because the spatial variation of the observed quantities is too large to be described by a single plane.

As a next step, we adopted an approach to fit the pixels by 2 or 3 planes so as to cover their large variation.
For the purpose, we chose two pixel groups by $\Delta N_{\mathrm{g}}(1)>0$ and $\Delta N_{\mathrm{g}}(1)\leq 0$, and three pixel groups with $\Delta N_{\mathrm{g}}(1) \geq 0.1$, $-0.1 < \Delta N_{\mathrm{g}}(1) <0.1$, and $\Delta N_{\mathrm{g}}(1) \leq -0.1$, for the-2 plane fit and 3-plane fit, respectively.
This grouping yielded a nearly equal number of pixels in each pixel group, and is not biased toward the spatial distribution of the individual pixels either on the shell or in the inner part.

The distributions of $\Delta N_{\mathrm{g}}(2)$ and $\widehat{N}_{\mathrm{g}}(2)$ and of $\Delta N_{\mathrm{g}}(3)$ and $\widehat{N}_{\mathrm{g}}(3)$ are shown in Figures \ref{fig:dNg_all_inner_shell_plot_error_map} and \ref{fig:dNg_plot_all_3region_dNgpm0p1_error_map}, respectively, where 2 and 3 in the brackets stand for the 2- and 3-plane fits.
We see that the distributions of $\Delta N_{\mathrm{g}}$ are similar to Figure \ref{fig:dNg_Nghat_all_plot_error_map}(a), showing a trend of decrease from the shell to the inner part.
Table \ref{tab:regression_results_0852} presents the coefficients $a$ and $b$.
The resultant hadronic and leptonic fractions (Table \ref{tab:estimate_hadron_and_lepton_0852}) have similar values within a small difference of $±6\%$ in the three plane fit.

Table \ref{tab:estimate_hadron_and_lepton_0852} also summarizes the values of $N_{\mathrm{g}}$ obtained by the present analysis.
Based on the present 3-plane results, we calculate the hadronic and leptonic gamma ray counts in each pixel, $\widehat{N}_{\mathrm{g}, i}^{\mathrm{hadronic}}$ and $\widehat{N}_{\mathrm{g}, i}^{\mathrm{leptonic}}$, as given by
\begin{equation}
    \widehat{N}_{\mathrm{g}, i}^{\mathrm{hadronic}}=a N_{\mathrm{p}, i}
\end{equation}
and
\begin{equation}
    \widehat{N}_{\mathrm{g}, i}^{\mathrm{leptonic}}=b N_{\mathrm{x}, i}
\end{equation}
for the three pixel groups.
All the errors are calculated in the same manner as \citetalias{2021ApJ...915...84F}, and are not repeated here.
Consequently, we find that the hadronic component occupies $(52 \pm 1)$\% of the total gamma ray count and the leptonic component $(48 \pm 1)$\%.
Further discussion is given in Section \ref{sec:discussion}.
Figure \ref{fig:hadronicgammafraction_all_kttv_div5_plotmap}(a) shows the hadronic fraction of gamma rays as a function of $N_{\mathrm{x}}$ for two ranges of $N_{\mathrm{p}}$ larger or smaller than $3.26\times 10^{21}$\,cm$^{-2}$.
We see a trend that the hadronic fraction decreases with $N_{\mathrm{x}}$ and increases with $N_{\mathrm{p}}$.
Figure \ref{fig:hadronicgammafraction_all_kttv_div5_plotmap}(b) shows the spatial distribution of the hadronic fraction.
We find that the fraction is increased in the inner part of the shell and decreased in the shell, in particular in the north and south where the X rays are enhanced.

\begin{figure*}[p]
\gridline{\fig{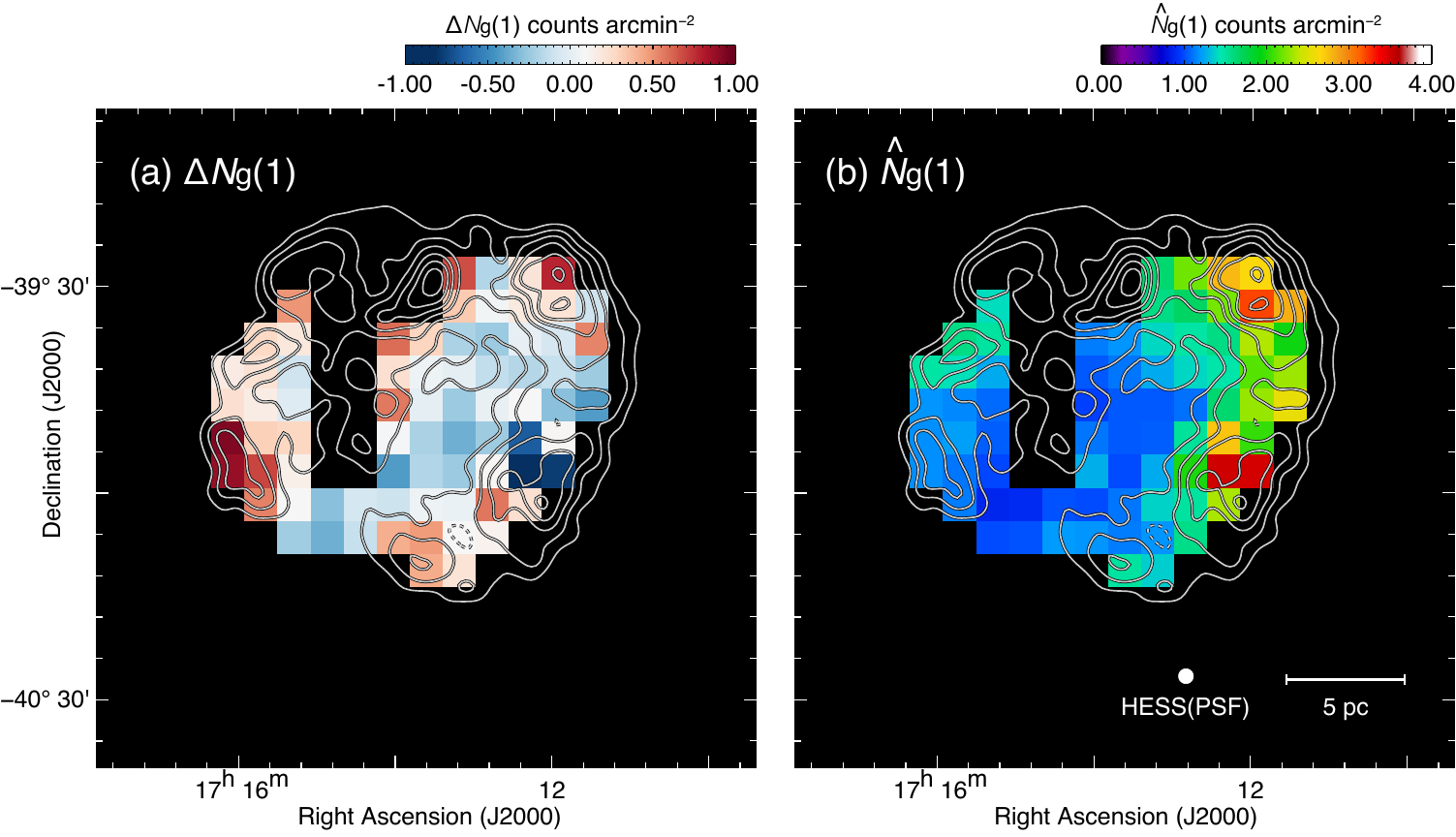}{.8\textwidth}{}}
\gridline{\fig{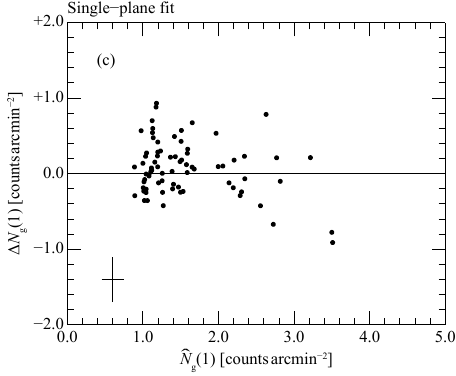}{0.5\textwidth}{}}
\caption{
Results of the single-plane fit for RXJ1713.
(a) Maps of $\Delta N_{\mathrm{g}}(1)$ derived from Equation (\ref{eqn:deltaNg}) and (b) $\widehat{N}_{\mathrm{g}}(1)$ derived from Equation (\ref{eqn:regression_model}).
The superposed contours in both Panels (a) and (b) indicate excess counts of TeV gamma rays.
The lowest contour level and contour intervals are 12 and 5 counts, respectively.
(c) A plot of $\Delta N_{\mathrm{g}}(1)$ with respect to $\widehat{N}_{\mathrm{g}}(1)$.
The horizontal- and vertical error bars shown in the left bottom corner indicate the median values of $\sigma(\widehat{N}_{\mathrm{g}}(1))$ and $\sigma(\Delta N_{\mathrm{g}}(1))$ \citepalias[see Equations (6) and (7) in][]{2021ApJ...915...84F}, respectively.
}
\label{fig:RXJ1713_dNg_Nghat_plotmap_all}
\end{figure*}%

\begin{figure*}
\gridline{\fig{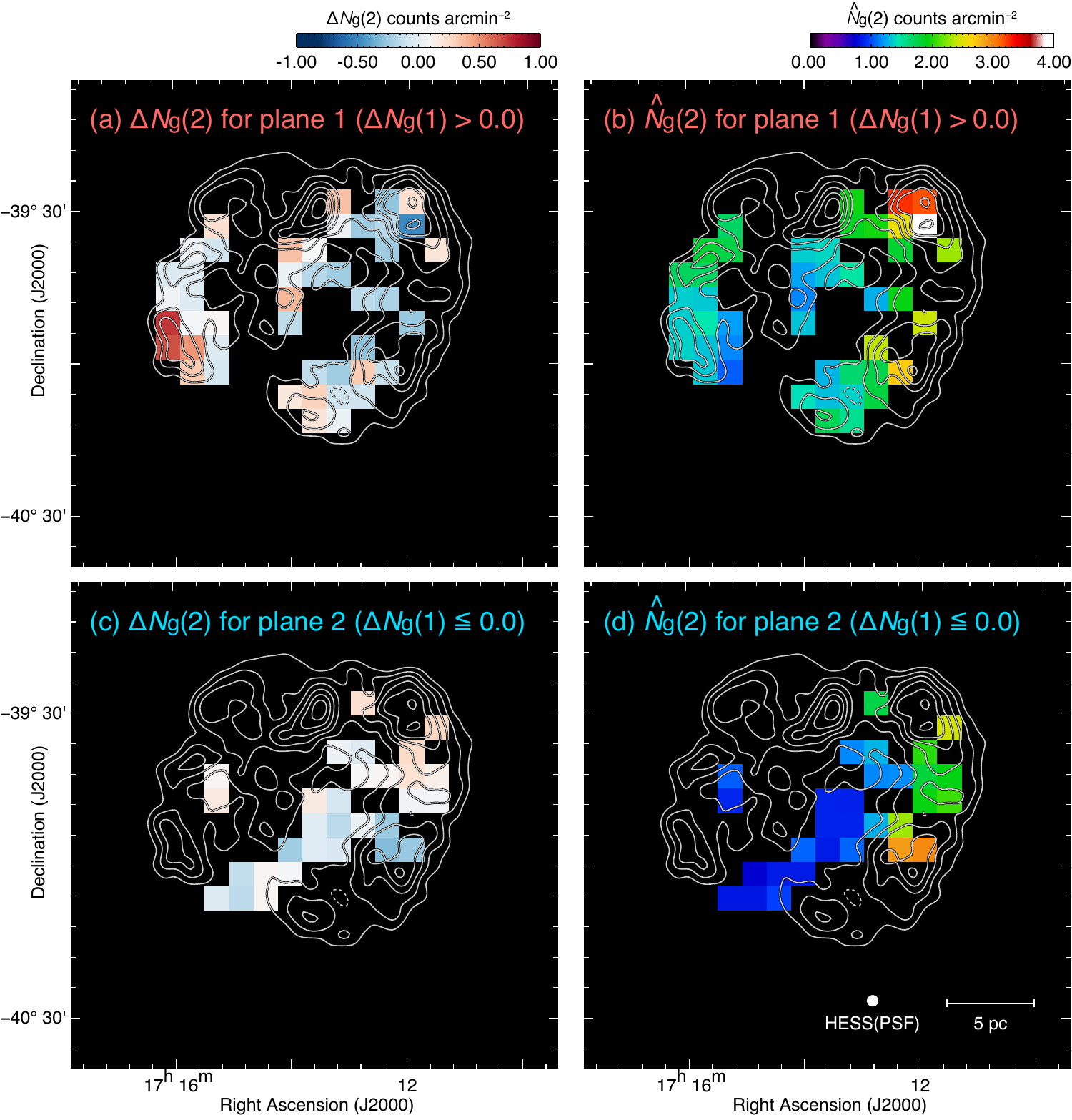}{.7\textwidth}{}}
\gridline{\fig{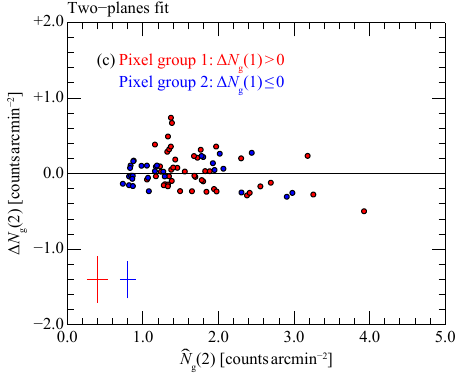}{0.5\textwidth}{}}
\caption{
Results of the two-planes fit for RXJ1713.
(a)--(d) Maps of $\Delta N_{\mathrm{g}}(2)$ (left panels) and $\widehat{N}_{\mathrm{g}}(2)$ (right panels) for the pixel groups 1 ($\Delta N_{\mathrm{g}}(1)>0$, top panels) and 2 ($\Delta N_{\mathrm{g}}(1)\le 0$, middle panels).
The superposed contours in each panel are identical to those in Figure \ref{fig:RXJ1713_dNg_Nghat_plotmap_all}.
(e) A plot of $\Delta N_{\mathrm{g}}(2)$ with respect to $\widehat{N}_{\mathrm{g}}(2)$.
The horizontal- and vertical error bars shown in the left bottom corner indicate the median values of $\sigma(\widehat{N}_{\mathrm{g}}(2))$ and $\sigma(\Delta N_{\mathrm{g}}(2))$, respectively.
The red circles and error bars are for the pixel group 1 and blue ones are for the pixel group 2.
}
\label{fig:RXJ1713_dNg_Nghat_all_dNggt0_plotmap}
\end{figure*}

\begin{figure}
\begin{center}
\includegraphics{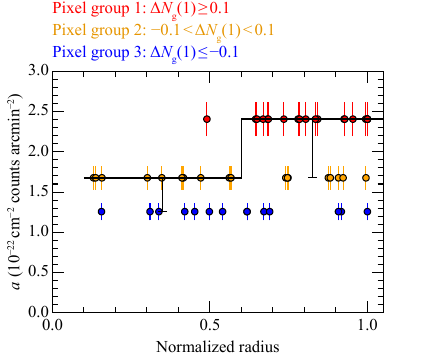}
\end{center}
\caption{
Fitted coefficient $a$ versus SNR radius for RXJ0852.
The error bar associated with each data point present standard error.
The red circles and error bars are for the pixel group 1, the orange and blue ones are for the pixel groups 2 and 3, respectively.
The median values of $a$ are shown for two bins of the normalized radius, with the vertical error bars showing the interquartile range (in each bin, the upper quartile is the same as the median)}.
\label{fig:ndist_vs_a_0852}
\end{figure}


\begin{deluxetable*}{lRRRRR}
\tablecaption{Summary of the Multiple Linear Regression for RXJ1713} \label{tab:regression_results_1713}
\tablewidth{0pt}
\tablehead{
\colhead{Model} & \colhead{$n$} & \colhead{$a$} & \colhead{$b$} & \colhead{$\chi ^{2} / \nu $} & \colhead{VIF}
}
\decimalcolnumbers
\startdata
Single-plane fit & 75 & 1.45 \pm 0.17 & 1.02\pm 0.24 & 1.74 & 2.0 \\
\hline
\multicolumn{5}{l}{Two-plane fit} \\
\hline
Shell ($\Delta N_{\mathrm{g}}(1)>0$) & 45 & 1.58 \pm 0.17 & 1.39\pm 0.24 & 0.85 & 1.6 \\
Inner part ($\Delta N_{\mathrm{g}}(1)\le 0$) & 30 & 1.09 \pm 0.12 & 1.01\pm 0.16 & 0.41 & 2.6 \\
\enddata
\tablecomments{
Columns (1): model name; (2): number of pixels of the data set; (3) and (4): estimated regression coefficients and their standard deviations; (5): reduced-$\chi^{2}$ (Equation (\ref{eqn:reduced-chi-square})); (6): VIF, less than 3 indicates that multicollinearity is not a significant issue in the analysis \citepalias[see Appendix C3 of][]{2021ApJ...915...84F}.
}
\end{deluxetable*}

\begin{deluxetable*}{lRRRRRRR}
\tablecaption{The hadronic and leptonic gamma-ray components in RXJ1713} \label{tab:estimate_hadron_and_lepton_1713}
\tablewidth{0pt}
\tablehead{ & & & \multicolumn2c{Hadronic component} & &  \multicolumn2c{Leptonic component} \\ 
\cline{4-5}
\cline{7-8}
\colhead{Model} & \colhead{$\langle\widehat{N}_{\mathrm{g}}\rangle$} & $\langle{N}_{\mathrm{p}}\rangle$ & \colhead{$\langle\widehat{N}_{\mathrm{g}}^{\mathrm{hadronic}}\rangle$} & \colhead{$\langle\widehat{N}_{\mathrm{g}}^{\mathrm{hadronic}}\rangle/\langle\widehat{N}_{\mathrm{g}}\rangle$} & \colhead{$\langle{N}_{\mathrm{x}}\rangle$} & \colhead{$\langle\widehat{N}_{\mathrm{g}}^{\mathrm{leptonic}}\rangle$} & \colhead{$\langle\widehat{N}_{\mathrm{g}}^{\mathrm{leptonic}}\rangle/\langle\widehat{N}_{\mathrm{g}}\rangle$} }
\decimalcolnumbers
\startdata
Single-plane fit & $1.56\pm 0.02$ & 0.72 & 1.04\pm 0.02 & (67\pm 1)\% & 0.50 & 0.51\pm 0.02 & (33\pm 1)\% \\
\hline
\multicolumn{8}{l}{Two-plane fit} \\
\hline
Whole & 1.60\pm 0.02 & 0.72 & 0.99\pm 0.01 & (62\pm 1)\% & 0.50 & 0.62\pm 0.01 & (38\pm 1)\%   \\
Shell ($\Delta N_{\mathrm{g}}(1)>0$) & $1.74\pm 0.03$ & 0.69 & 1.09\pm 0.02 & (63\pm 1)\% & 0.47 & 0.65\pm 0.02 & (37\pm 1)\%  \\
Inner part ($\Delta N_{\mathrm{g}}(1)\le 0$) & 1.39\pm 0.03 & 0.76 & 0.83\pm 0.02 & (60\pm 2)\% & 0.55 & 0.56\pm 0.02 & (40\pm 2)\% \\
\enddata
\tablecomments{
Columns (1): model name; (2), (3) and (6): spatial averages of observed $N_{\mathrm{g}}$ (counts\,arcmin$^{-2}$), $N_{\mathrm{p}}$ ($10^{22}$\,cm$^{-2}$) and $N_{\mathrm{x}}$ ($10^{2}$\,counts\,s$^{-1}$\,degree$^{-2}$); (4): spatial average of the predicted hadronic-origin gamma-rays (counts\,arcmin$^{-2}$) given by $\sum_{i}^{n}(\widehat{N}_{\mathrm{g}, i}^{\mathrm{hadronic}})/n$ and its standard deviation; (5): fraction of the hadronic component; (7) and (8): same as (4) and (5) but for the predicted leptonic-origin gamma-rays.
}
\end{deluxetable*}


\subsection{Fitting of the Hadronic and Leptonic Components in RXJ1713}
Figure \ref{fig:RXJ1713_dNg_Nghat_plotmap_all} reproduces the results of the single-plane fit from \citetalias{2021ApJ...915...84F}.
Following the multi-plane fit in RXJ0852, we applied a two-plane fit to the data pixels of RXJ1713 for the same dataset in \citetalias{2021ApJ...915...84F}.
In the present fit, the pixels were grouped into two with $\Delta N_{\mathrm{g}}(1)>0$ and $\Delta N_{\mathrm{g}}(1)<0$.
The fit results are listed in Figure \ref{fig:RXJ1713_dNg_Nghat_all_dNggt0_plotmap} and Table \ref{tab:regression_results_1713}.
The reduced-$\chi^{2}$ in Table \ref{tab:regression_results_1713} shows that the values in the 2-plane fit are decreased further to 0.41--0.85 as compared with that 1.74 in the single plane fit.
We find the present results show that the fit results have smaller errors.
The distributions of $\Delta N_{\mathrm{g}}(2)$ of the two pixel groups are shown in Figures \ref{fig:RXJ1713_dNg_Nghat_all_dNggt0_plotmap}(a)--(d).
We find that the $\Delta N_{\mathrm{g}}(1)>0$ pixel group and the $\Delta N_{\mathrm{g}}(1)\leq 0$ pixel group are distributed both in the inner part and the shell part.
Table \ref{tab:estimate_hadron_and_lepton_1713} shows the hadronic and leptonic gamma ray components.
Consequently, we obtain that the hadronic component occupies $(62 \pm 5)$\% of the total gamma ray count and the leptonic component $(38 \pm 5)$\% in RXJ1713.
These indicate that the hadronic gamma rays are by 5\% weaker within systematic erros of \citetalias{2021ApJ...915...84F}, while the basic trend remains the same.

\section{Discussion}\label{sec:discussion}
\subsection{Hadronic/Leptonic Gamma Ray Ratio}
\citetalias{2012ApJ...746...82F} showed that the ISM proton distribution corresponds well to the gamma ray distribution, whereas \citetalias{2012ApJ...746...82F} did not exclude the leptonic components.
If we assume an extreme case where the leptonic gamma ray counts are by more than an order of magnitude higher than the hadronic gamma rays everywhere in an SNR, the hadronic components become almost completely masked by the leptonic components.
Then, it is very difficult to quantify the hadronic components from observations.
However, the present work along with \citetalias{2021ApJ...915...84F} revealed that the hadronic components and the leptonic components are comparable in the gamma ray counts and competing with each other in the two SNRs RXJ1713 and RXJ0852.
A probable reason for the situation is the large mass of the target interstellar protons which enhances the hadronic gamma rays.
The two SNRs are of the core-collapse type, which are naturally associated with massive neutral gas either in atomic or molecular form.
On the other hand, in Type Ia SNRs such neutral gas may not be massive enough to produce strong hadronic gamma rays, leading to insignificant hadronic components.

\subsection{Quantification of the CR Energy}
Quantification of the CR energy density in an SNR is crucial in verifying the acceleration of CRs.
The present work along with \citetalias{2021ApJ...915...84F} demonstrates that the present methodology is able to quantify the hadronic gamma rays and the CR proton energy $W_{\mathrm{p}}$ in RXJ0852 and RXJ1713.
The energy is calculated based on the associated interstellar proton density along with the gamma ray luminosity.
The CR proton energy $W_{\mathrm{p}}$ in an SNR is expressed as follows;
\begin{equation}
W_{\mathrm{p}} = t_{\mathrm{pp}} L_{\mathrm{g}} f,
\end{equation}
where $t_{\mathrm{pp}}$ is the cooling time of CR protons and $L_{\mathrm{g}}$ the total gamma ray luminosity \citep{2006A&A...449..223A}, and $f$ is the hadronic fraction.
The total CR proton energy $W_{\mathrm{p}}$ above 1\,GeV in RXJ0852 is given as follows \citep{2018A&A...612A...7H};
\begin{equation}
    W_{\mathrm{p}} = (7.1 \pm 0.3_{\mathrm{stat}} \pm 1.9_{\mathrm{syst}}) 10^{49}\left(\frac{d}{750\,\mathrm{pc}}\right)^{2}\left(\frac{n_{\mathrm{p}}(\mathrm{ISM})}{1\,\mathrm{cm}^{-3}}\right)^{-1}\,\mathrm{erg},
\end{equation}
where $d$ is the distance, 750\,pc, and $n_{\mathrm{p}}(\mathrm{ISM})$ is the ISM proton density, $\sim 90$\,cm$^{-3}$, by adopting the shell radius $\sim 11.8$\,pc and the thickness $\sim 5.6$\,pc.
$W_{\mathrm{p}}$ is estimated to $W_{\mathrm{p}} = 7.7\times 10^{47}$\,erg by adopting $n = 90$\,cm$^{-3}$, where the hadronic gamma rays are assumed to be 100\% of the total gamma ray counts.
The present work has shown that a fraction of the hadronic gamma ray counts is 52\% in the whole SNR, and $W_{\mathrm{p}}=4.0\times 10^{47}$\,erg.

In RXJ1713 $W_{\mathrm{p}}$ above 1\,GeV is calculated by the relationship 
\begin{equation}
    W_{\mathrm{p}}= (1 \mbox{--} 3)\times 10^{50}\left(\frac{d}{1\,\mathrm{kpc}}\right)^{2}\left(\frac{n_{\mathrm{p}}(\mathrm{ISM})}{1\,\mathrm{cm}^{-3}}\right)^{-1}\,\mathrm{erg},
\end{equation}
\citep{2006A&A...449..223A}.
We then obtain $W_{\mathrm{p}}=(0.5\mbox{--}1.4)\times 10^{48}$\,erg for $n=130$\,cm$^{-3}$ \citepalias{2012ApJ...746...82F,2021ApJ...915...84F} and a hadronic fraction of the present work 0.62.
This energy is similar to that of RXJ0852.

We calculate roughly the energy density of CR protons in the two SNRs.
In RXJ0852, we calculate the energy density to be $2.3\times 10^{-12}$\,erg\,cm$^{-3}$ through dividing $W_{\mathrm{p}}$ by the volume of the SNR shell having an outer radius of 11.8\,pc and an inner radius of 6.2\,pc.
We also apply the same method to RXJ1713 which is modeled by a shell having an outer radius of 10.1\,pc and an inner radius of 5.9\,pc \citepalias{2012ApJ...746...82F} and obtained the average density to be $(3.1\mbox{--}8.6)\times 10^{-12}$\,erg\,cm$^{-3}$.
The CR energy densities above 1\,GeV in the two SNRs are inferred to be $1\mbox{--}5$\,eV\,cm$^{-3}$, which gives a secure lower limit for the energy density.
This value is significantly greater than the energy density of the CR sea $\sim 1$\,eV\,cm$^{-3}$ in the Galaxy which is peaked at $\sim 100$\,MeV if we take into account the volume filling factor of the ISM protons, $\sim 0.1$ (see below in this subsection).

Here we note that the numbers of the pixels used in the present fits were reduced by $\sim 70$\% from the original datasets, because of the exclusion of the pixels with contamination by the PWN (RXJ0852) and with inaccurate ISM density (RXJ1713), as well as those outside the gamma ray shells.
In the above we did not include the corrections.
We suggest the energy values above may be reduced to be $\sim 70$\% if they are taken into account.

The efficiency of the gamma ray emission via IC scattering by CR electrons is higher than that of the CR protons by two orders of magnitude \citepalias[e.g.,][]{2010ApJ...708..965Z}. The leptonic gamma rays then become comparable to the hadronic gamma rays, even if the electrons are $\sim 1/100$ of the protons in number.
The CR electron energy density is therefore likely in the order of $10^{45}$\,erg, a negligibly small fraction of the total CR energy density.

The CR proton energy $W_{\mathrm{p}}$ calculated in the two SNRs $(3\mbox{--} 9)\times 10^{47}$\,erg is smaller than the value $\sim 10^{50}$\,erg, which is an expected $W_{\mathrm{p}}$ often quoted if SNRs are the major source of CRs in the Galaxy.
This seems to be a difficulty of the SNR origin of the Galactic CRs.
There are however two factors which need to be taken into account.
One is a volume filling factor of the target interstellar protons.
The filling factor is roughly estimated to be $\sim 0.1$ because the dense gas is highly clumped within the shell in RXJ1713 \citepalias{2012ApJ...746...82F} as well as in RXJ0852 \citepalias{2021ApJ...915...84F}.
It is thus possible that the current $W_{\mathrm{p}}$ in an SNR is by an order of magnitude smaller than $(3\mbox{--}9)\times 10^{47}$\,erg derived in the present work, because the majority of the CR protons are not contributing to $W_{\mathrm{p}}$ via the p-p reaction.
The other is the escaping CRs from the SNR shell which lowers $W_{\mathrm{p}}$ by a factor of $\gtrsim 10$ over an SNR age of $>10^{5}$\,yr \citep{2007ApJ...665L.131G} within an SNR boundary, while an estimate of such CRs yet involves large uncertainty (\citealt{2012ApJ...749L..35U}, W44; \citealt{2022ApJ...938...94A}, Puppis~A; see also \citealt{2018A&A...612A...6H}).
Accordingly, we argue that SNRs as the most promising sources of CRs in the Galaxy is a plausible picture. The present work as well as \citetalias{2017ApJ...850...71F} provides the best quantification of the CR proton energy up to 100\,TeV in the two representative young SNRs that seem to be most active in particle acceleration in the solar vicinity.

A scatter plot between a SNR age and CR proton energy $W_{\mathrm{p}}$ is presented for eleven SNRs by \citet{2021ApJ...919..123S}.
The plot shows a trend that $W_{\mathrm{p}}$ increases in the first several times $10^{3}$-yr of a young SNR from $10^{48}$\,erg up to $3\times 10^{49}$\,erg.
The phase is followed by decreasing $W_{\mathrm{p}}$ probably due to CR escape, which corresponds to the middle-aged SNRs.
The trend suggests the total CR proton energy as large as $10^{50}$\,erg over the SNR lifetime is not unrealistic by considering the significant CR escape.

\subsection{Possible Variation of the CR Energy Density between the Shell and the Inner Part}
The fit by 2--3 planes instead of a single plane was required to accommodate the strong variation of the high energy radiation of RXJ0852, in particular, the steep increase of the X rays toward the shell.
The multiplane fit allows some variation of the coefficients $a$ and $b$, which tend to increase from the inner part to the shell part.
A similar trend is also found in RXJ1713, suggesting that the variation may be common in young shell type young SNRs.

In RXJ0852, we derived the variation of $a$ by assuming spherical symmetry of the SNR.
For the normalization, first, separation of each pixel from the central position $(\alpha_{\mathrm{J2000}}, \delta_{\mathrm{J2000}}) = (-133\fdg 0, -46\fdg 37)$ \citep{2018A&A...612A...6H} was measured.
Then, the SNR was divided into 12 fan-shaped segments whose apex has an opening angle of 30 degrees at the center.
In each segment the maximum separation of pixels was chosen as the radius.
The separation of a pixel was divided by the radius and determined as the normalized radius of the pixel.
Figure \ref{fig:ndist_vs_a_0852} shows a plot of $a$ as a function of the normalized radius.
We find a trend that $a$ becomes enhanced by $\sim 30$\% on average at normalized radii in the shell at a normalized radius 0.6--1.0 than the inner part at a normalized radius less than 0.6.
In RXJ1713, we performed the same analysis, but found no trend like in RXJ0852, because the pixels fit by the two planes are more mixed up spatially than in RXJ0852, resulting in a thicker shell.

The coefficient $a$ represents the p-p reaction coefficient and the CR proton energy density.
The coefficient $b$ depends on the IC scattering, the CMB density and inversely on the B field energy density ($B^{2}$), as well as the CR electron energy spectral index for converting the nonthermal X rays into the gamma ray photons.   

The variation of $a$ therefore may mean an increase of the CR energy density from the inner part to the shell by $\sim 40$\%, which may be in accord with the enhanced CR acceleration in the shell where the shock speed is higher as shown by the theoretical works \citepalias[see Figure 2 of][]{2010ApJ...708..965Z}.
The trend in $b$ may be more complicated than in $a$, since $b$ depends on the CR electron spectrum as well as the magnetic field strength.
For the electron spectrum in RXJ1713, \citet{2015ApJ...799..175S} showed that the electron spectral index becomes harder in the shell than in the inner part, which may possibly increase $b$ in the shell.
On the other hand, the magnetic field strength increases in the shell due to the turbulent amplification, causing the decrease of $b$ near the CO clumps \citep{2012ApJ...744...71I}.
So, more details of the trends especially in $b$ remain as a future issue when more reliable variations of $a$ and $b$ become available.
At any rate, in the CR energetics of the SNR, $a$ plays a major role as compared with $b$.

\subsection{The Origin of $N_{\mathrm{p}}$ and $N_{\mathrm{x}}$}
In the present methodology, the significant difference in the distribution between $N_{\mathrm{p}}$ and $N_{\mathrm{x}}$ is essential in quantifying the two components.
It is impossible to disentangle the two components reliably if the two distributions are very similar with a correlation coefficient close to 1.0.
The difference is naturally produced by the evolutionary process of $N_{\mathrm{p}}$ and $N_{\mathrm{x}}$, and can be expected to hold in the other SNRs including HESSJ1731$-$347 and RCW~86. 
The $N_{\mathrm{p}}$ distribution with a central cavity is produced by the compression of the surrounding clumpy interstellar medium by the stellar winds of the progenitor high mass star.
The time scale of the compression is in the order of $\sim 10$\,Myr.
On the other hand, the $N_{\mathrm{x}}$ distribution, a combined result of the accelerated CR electrons and the magnetic field, is produced via particle acceleration and shock-cloud interaction within 1000\,yr after the supernova explosion; the shock front driven by the explosion accelerates particles over the few 1000\,yrs and amplifies the turbulence and the magnetic field around the dense clumps via shock-cloud interaction in the recent several 100\,yrs in the present two SNRs \citep{2010ApJ...724...59S,2012ApJ...744...71I}.
The physical processes responsible for the $N_{\mathrm{p}}$ and $N_{\mathrm{x}}$ distributions have therefore different origins and timescales by orders of magnitude, although some correlation between them is expected.

The reduced $N_{\mathrm{g}}$ and $N_{\mathrm{p}}$ correlation in RXJ0825 vs RXJ1713 (Figure \ref{tab:regression_results_0852}(b)) is probably because of the small mass of the CO clumps in RXJ0852, $\sim 1000$\,$M_{\sun}$, which is about 10\% of that in RXJ1713 \citepalias{2017ApJ...850...71F}.
The \ion{H}{1} mass is $\sim 10^{4}$\,$M_{\sun}$ in the two SNRs \citepalias{2012ApJ...746...82F,2021ApJ...915...84F}.
The small CO mass results in less coupling between $N_{\mathrm{p}}$ and $N_{\mathrm{x}}$ due to less shock cloud interaction.
\citet{2013ApJ...778...59S} showed that the non-thermal X rays associated with the CO clumps are nearly proportional to the CO clump mass in RXJ1713 \citep[see Figure~11 of ][]{2013ApJ...778...59S}.
The CO mass is the condition provided prior to the supernova.
A probable reason for the small CO clump mass is a smaller $\mathrm{H}_{2}/\mbox{\ion{H}{1}}$ ratio around RXJ0852, which is in the outer solar circle at a high latitude \citepalias{2017ApJ...850...71F}.

\subsection{Future Prospects}
The present methodology has a potential to be extended to the other young SNRs, where the distributions of $N_{\mathrm{p}}$ and $N_{\mathrm{x}}$ are available.
$N_{\mathrm{x}}$ is the nonthermal X rays which is emitted by CR electrons in the same energy range with the gamma ray production via the IC scattering.
It is however not often the case that the nonthermal X rays of SNRs are imaged with sufficiently high quality.
We expect a few TeV gamma ray SNRs including HESS~J1731 and RCW~86 are promising candidates in future for which the present methodology is applicable.
Currently, the angular resolution of the gamma ray observations is not high enough for them because the number of independent pixels is limited to only $\sim 10$ due to their small diameter of $\sim 0\fdg 5$.
In this context, we point out that simple diagnostics by using two scatter plots of $N_{\mathrm{g}}$ with $N_{\mathrm{p}}$ and $N_{\mathrm{x}}$ like those in Figures \ref{fig:NgNxNp_all_plot_error} and \ref{fig:RXJ1713_NgNxNp_all_plot_error} provide a useful test toward the quantification of the gamma ray components.
The two scatter plots of RXJ0852 and RXJ1713 show good correlations among them, which are consistent with the significant contribution to $N_{\mathrm{g}}$ of both $N_{\mathrm{p}}$ and $N_{\mathrm{x}}$, and represent essential features of the gamma ray origin.

Another possibility in future is toward the middle-aged SNRs.
The observations of the interstellar protons have been extensively made already for tens of the middle-aged SNRs (e.g., \citealt{2013ApJ...768..179Y}, W44; \citealt{2008A&A...481..401A}, W28; \citealt{2017AIPC.1792d0039Y}, IC443; \citealt{2021ApJ...923...15S}, G346.6$-$0.2; \citealt{2018ApJ...864..161K}, Kes~79; \citealt{2022ApJ...938...94A}, Puppis~A).
Their X rays are mostly of thermal origin and do not provide a measure of the CR electrons.
Non-thermal radio emission may be used instead of the X rays, while the energy range is not identical to the gamma ray emitting electrons, causing uncertainty depending on the energy spectrum of the CR electrons.
Nonetheless, it might be worth testing how the methodology works in the middle aged SNRs.
Another uncertainty here is the estimate of the mass of target interstellar protons.
Very often, the target protons are only roughly assumed to be 1\,cm$^{-3}$ in these SNRs in the literature \citep[e.g.,][]{2012ApJ...744...39E}.
Since the ISM density is a critical quantity in the CR energetics, accurate estimates of the ISM density are needed.
In addition, the low spatial resolution of the $<10$\,GeV gamma-ray emission does not allow us to extend easily our methodology to the middle aged SNRs at present. 

\section{Conclusions}\label{sec:conclusions}
We have carried out quantification of the hadronic and leptonic gamma rays in the SNR RXJ0852 by employing the methodology which expresses the hadronic and leptonic gamma ray components by linear combinations of $N_{\mathrm{p}}$ and $N_{\mathrm{x}}$ under an assumption of uniform CR energy density and magnetic field \citepalias{2021ApJ...915...84F}.
The methodology was applied to the gamma ray datasets of RXJ0852 obtained by H.E.S.S., which are combined with the \textit{Suzaku} X ray data and the NANTEN/ATCA-Parkes ISM data.
A new feature was to use multi-plane fit instead of the single plane fit, which allowed to accommodate the steep spatial variation of X rays by somewhat relaxing the assumptions of the uniform parameters.
\begin{enumerate}
    \item The present quantification employed fitting by two/three flat planes in a 3D space subtended by $N_{\mathrm{p}}$-$N_{\mathrm{x}}$-$N_{\mathrm{g}}$.
    Along with RXJ0852, RXJ1713 was re-examined under the modification of a multi-plane fit to \citetalias{2021ApJ...915...84F}.
    The multi-plane fit was required to accommodate the steep change of the X rays etc, in the shell of RXJ0852 and was able to reduce the fit errors in the two SNRs, without significantly altering the assumptions of uniform CR energy and magnetic field strength within the SNR.
    \item The results show that the gamma rays of RX J0852 consist of both hadronic gamma ray components: $\mbox{the leptonic gamma ray components} = (52\pm 1) : (48 \pm 1)$\% in $N_{\mathrm{g}}$.
    This indicates that the two gamma ray components are in the same order of magnitude, and are competing with each other in the SNR.
    This provides a second case next to RXJ1713, where the hadronic gamma rays are significant in an SNR, proving the cosmic ray origin in a young SNR.
    \item The CR proton energy is calculated to be $(3\mbox{--}9)\times 10^{47}$\,erg in both RXJ0852 and RXJ1713, about a half or one third of that calculated without the component quantification \citepalias{2017ApJ...850...71F,2021ApJ...915...84F}.
    The energy is still compatible with a scheme that SNRs are the dominant source of the Galactic CRs, if we consider the volume filling factor of the target protons in the order of 0.1 and the large energy carried out by escaping CRs.
    So, the CR energy stored in an SNR at a given instance is always significantly smaller than $10^{50}$\,ergs, while the energy is provided eventually if the energy were integrated in space and time over the whole lifetime of an SNR.
    The CR electron energy is likely smaller by two orders of magnitude than the CR proton energy, where the exact value depends on the magnetic field not directly measured, as already shown by the theoretical works.
    \item As a consequence of the multi-plane fit, we find in RXJ0852 that there is a trend of increase of the CR energy density in the shell as compared with the inner part of shell, which seems to be consistent with the theoretical models of CR acceleration.
    \item The present work as well as \citetalias{2021ApJ...915...84F} has demonstrated that the spatial distribution of the interstellar protons and nonthermal X rays [as a proxy of the CR electrons] are essential in quantifying the hadronic and leptonic gamma rays.
    The gamma ray spectral fit is not able to achieve such quantification due to large freedom in fit by tuning the physical parameters for either of the hadronic and leptonic models.
    The present methodology will be applicable to other TeV gamma SNRs, if the gamma ray resolution becomes higher by more than a factor of 2 \citep[e.g., by Cherenkov Telescope Array,][]{2019scta.book.....C}.
    In the middle-aged GeV gamma ray SNRs, such quantification is still not in the course of application.
    The low resolution in the GeV gamma rays and the lack/weakness of the nonthermal X rays do not allow to extend easily the methodology to the middle aged SNRs at present. 
\end{enumerate}




\begin{acknowledgments}
The NANTEN project is based on a mutual agreement between Nagoya University and the Carnegie Institution of Washington (CIW). We greatly appreciate the hospitality of all the staff members of the Las Campanas Observatory of CIW. We are thankful to many Japanese public donors and companies who contributed to the realization of the project.
This work is financially supported by a grant-in-aid for Scientific Research (KAKENHI, Nos.\ JP20H01945, JP21H0040, and JP22H00152) from Japan Society for the Promotion of Science (JSPS).
\end{acknowledgments}

\vspace{5mm}
\facilities{H.E.S.S.}

\software{HEASoft \citep{2014ascl.soft08004N}}

\appendix
\begin{figure*}[h]
\begin{center}
\includegraphics[width=\linewidth]{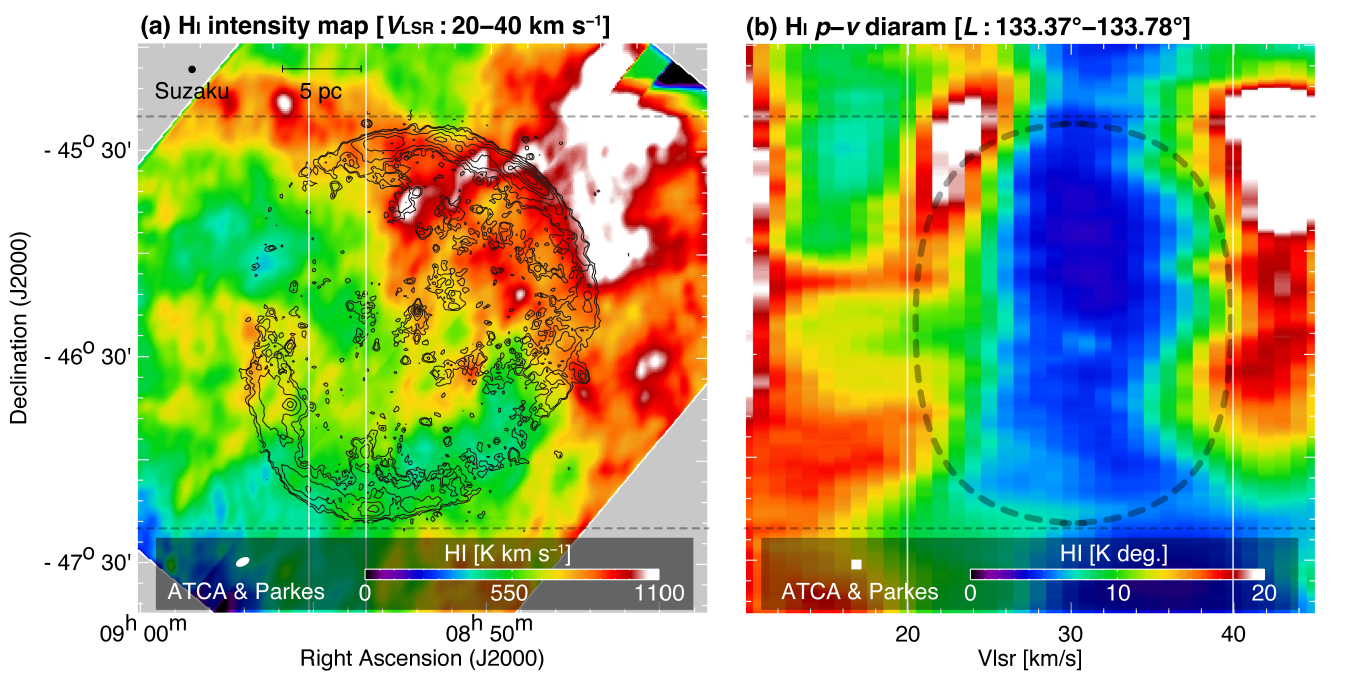}
\caption{(a) Integrated intensity map of \ion{H}{1} for RXJ0852.
The integration range is from 20 to 40\,km\,s$^{-1}$.
The superposed contours are the X-rays, and the contour levels are 2.0, 2.5, 4.0, 6.5, 10.0, 14.5, and $20.0\times 10^{-4}$\,counts\,s$^{-1}$\,pixel$^{-1}$.
(b) Position–velocity (p-v) diagrams of \ion{H}{1}.
The integration range is 133\fdg37 to 133\fdg78 in the R.A. for p-v diagram.
}
\label{fig12}
\end{center}
\end{figure*}%

\section{Velocity Range of the Associated ISM in RXJ0852}\label{sec:associated_ISM}
The distribution of the interstellar protons is derived based on the observational data of the interstellar hydrogen including both atomic and molecular hydrogen by \citetalias{2017ApJ...850...71F}.
In \citetalias{2017ApJ...850...71F}, a wide velocity range from 0 to 50\,km\,s$^{-1}$ in $V_{\mathrm{LSR}}$ was used.
In the present work, we chose a moderate range 20--40\,km\,s$^{-1}$.
This is motivated by the typical velocity ranges of the interacting ISM are $\sim 20$\,km\,s$^{-1}$ in several well studied SNRs \citep[e.g.,][]{2012ApJ...746...82F,2014ApJ...788...94F,2021ApJ...923...15S,2021ApJ...919..123S}.
Figure \ref{fig12} shows the present range in a position-velocity diagram, indicating the shell locus with expansion. The net difference of the adopted velocity range from that in \citetalias{2017ApJ...850...71F} is small less than 40\% in the $N_{\mathrm{g}}$ quantification as confirmed by comparing the results for the quantifications for the two velocity ranges.

\begin{figure*}
\begin{center}
\includegraphics[width=\linewidth]{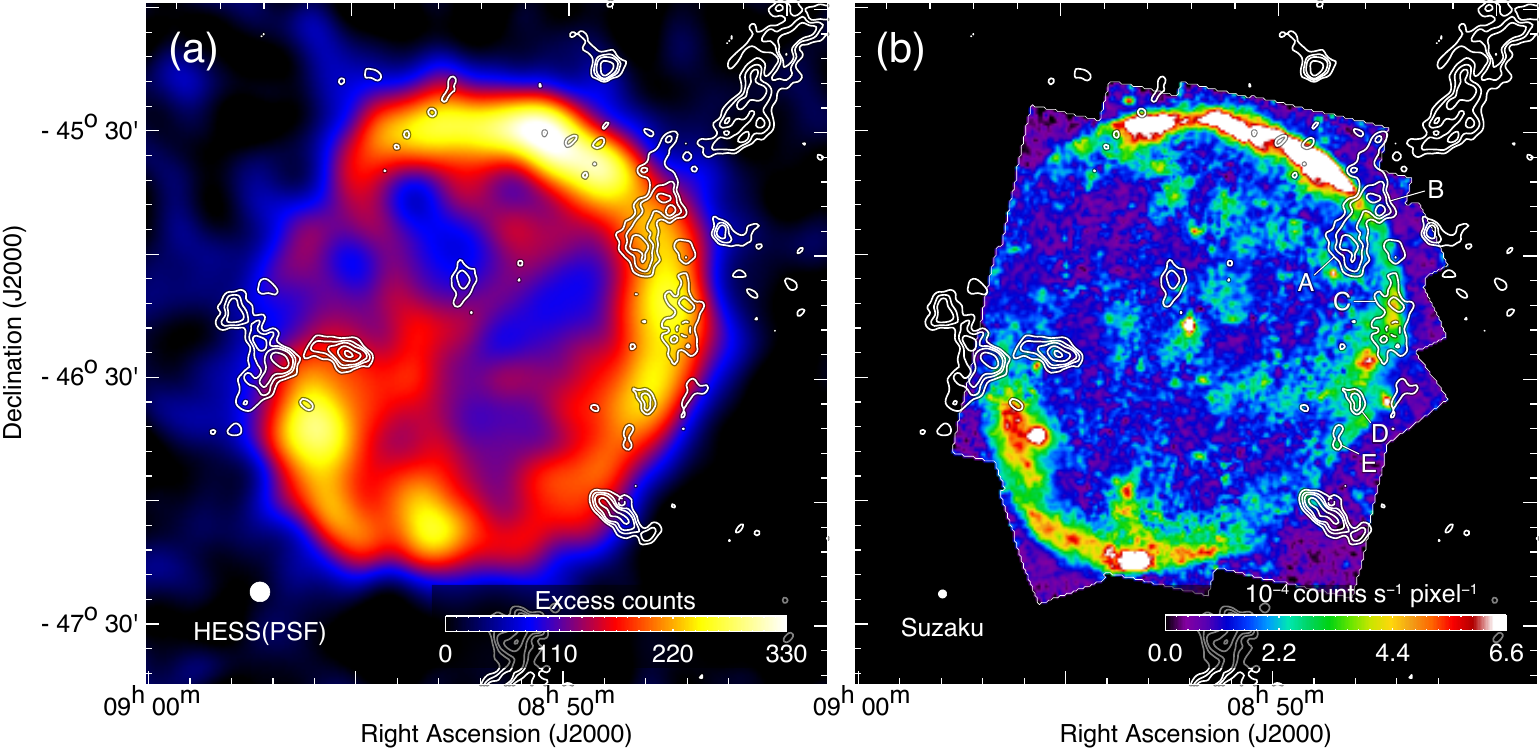}
\end{center}
\caption{
Maps of (a) TeV gamma ray excess counts ($E>100$\,GeV) \citep{2018A&A...612A...7H} and (b) X ray counts ($E=2.0$--5.7\,keV) \citepalias{2017ApJ...850...71F}.
The superposed contours indicate $^{12}$CO($J = 1\mbox{--}0$) intensity at $V_{\mathrm{LSR}}=20$--40\,km\,s$^{-1}$ whose contour levels are 2.5, 5.0, 7.5, 12.5, and 17.5\,K\,km\,s$^{-1}$.
Some molecular clouds in the western part of the shell are indicated as A--E.
}\label{fig:RXJ0852_appendix_f01}
\end{figure*}

\section{Spatial Comparisons with CO in RXJ0852}
Figure \ref{fig:RXJ0852_appendix_f01} shows a spatial comparison among the TeV gamma-rays, X-rays, and CO in RXJ0852.
We can find that CO clouds are distributed inside or on the gamma-ray and
X-ray shells.
It is noteworthy that some molecular clouds, especially for A, have clear anticorrelations with X-rays, suggesting that shock-cloud interactions with the magnetic amplification occurred.
In other words, these clouds are embedded within the SNR shell and act as targets for the accelerated cosmic-ray protons.

\bibliography{test}{}


\end{document}